\documentclass[fleqn,usenatbib]{mnras}

\usepackage{pdflscape}

\usepackage[T1]{fontenc}
\usepackage{ae,aecompl}

\usepackage{graphicx}	
\usepackage{amsmath}	
\usepackage{amssymb}	

\usepackage{adjustbox,lipsum} 

\title[X-ray sources in 47 Tuc]{\textit{Chandra} studies of the globular cluster 47 Tucanae: A deeper X-ray source catalogue, five new X-ray counterparts to millisecond radio pulsars, and new constraints to r-mode instability window
}

\author[Bhattacharya et al.]{
Souradeep~Bhattacharya$^{1,2}$\thanks{E-mail: f2012553@pilani.bits-pilani.ac.in},
Craig O.~Heinke$^{2}$,
Andrey I.~Chugunov$^{3}$,
\newauthor ~Paulo C.~C.~Freire$^{4}$,
Alessandro~Ridolfi$^{4}$
and Slavko~Bogdanov$^{5}$
\\
\\
$^{1}$Birla Institute of Technology and Science, Pilani, Rajasthan-333031, India\\
$^{2}$Physics Department, University of Alberta, 4-183 CCIS, Edmonton, AB T6G 2G7, Canada\\
$^{3}$Ioffe Institute, Polytekhnicheskaya 26, 194021 St. Petersburg, Russia\\
$^{4}$Max-Planck-Institut f\"ur Radioastronomie, Auf dem H\"ugel 69, D-53121 Bonn, Germany\\
$^{5}$Columbia Astrophysics Laboratory, Columbia University, 550 West 120th Street, New York, NY 10027, USA\\
}

\date{Accepted XXX. Received YYY; in original form ZZZ}

\pubyear{2017}

\begin{document}
\label{firstpage}
\pagerange{\pageref{firstpage}--\pageref{lastpage}}
\maketitle

\begin{abstract}

We combined \textit{Chandra} ACIS observations of the globular cluster 47 Tucanae (hereafter, 47 Tuc) from 2000, 2002, and 2014-15 to create a deeper X-ray source list, and study some of the faint radio millisecond pulsars (MSPs) present in this cluster. We have detected 370 X-ray sources within the half-mass radius (2$'$.79) of the cluster, 81 of which are newly identified, by including new data and using improved source detection techniques. The majority of the newly identified sources are in the crowded core region, indicating cluster membership. We associate five of the new X-ray sources with chromospherically active BY Dra or W UMa variables identified by Albrow et al. (2001). We present alternative positions derived from two methods, centroiding and image reconstruction, for faint, crowded sources. We are able to extract X-ray spectra of the recently discovered MSPs 47 Tuc aa, 47 Tuc ab, the newly timed MSP 47 Tuc Z, and the newly resolved MSPs 47 Tuc S and 47 Tuc F. Generally, they are well fit by black body or neutron star atmosphere models, with temperatures, luminosities and emitting radii similar to those of other known MSPs in 47 Tuc, though 47 Tuc aa and 47 Tuc ab reach lower X-ray luminosities. 
We limit X-ray emission from the full surface of the rapidly spinning (542 Hz) MSP 47 Tuc aa, and use this limit to 
put an upper bound for amplitude
of r-mode oscillations in this pulsar as $\alpha<2.5\times 10^{-9}$ and  constrain the shape of the r-mode instability window.

\end{abstract}

\begin{keywords}
stars: neutron--pulsars: general--globular clusters: individual: NGC 104
\end{keywords}


\section{Introduction}

In the cores of globular clusters (GCs), the very high stellar densities result in an extremely high rate of dynamical interactions. These produce a large number and variety of close binary systems, many of them unique to this environment, including low-mass X-ray binaries (LMXBs; \citealt{Clark 1975}), radio millisecond pulsars (MSPs; \citealt{Johnston 1992}), X-ray active binaries (ABs; \citealt{Bailyn et al. 1990}; \citealt{Dempsey et al. 1993}; \citeauthor{Grindlay et al. 2001a} \citeyear{Grindlay et al. 2001a}), and cataclysmic variables (CVs; \citeauthor{Pooley et al. 2003} \citeyear{Pooley et al. 2003}), which can engage in mass transfer. Such compact binaries in globular clusters have been effectively discovered by recent X-ray studies, especially using the \textit{Chandra X-Ray Observatory} (reviews include  \citeauthor{Verbunt Lewin 2004} \citeyear{Verbunt Lewin 2004}, and \citeauthor{Heinke 2010} \citeyear{Heinke 2010}). Detailed analyses of such sources have also made it possible to probe deeper into the formation and evolution of such remarkable objects \citep{Ivanova et al. 2006, Ivanova et al. 2008}.

MSPs are thought to have been produced from LMXBs when a low-mass companion star spins up the neutron star (NS) to millisecond periods by transferring angular momentum \citep{Bhattacharya van den Heuvel 1991, Papitto et al. 2013}. MSPs can produce X-ray radiation of both thermal origin--blackbody-like radiation from a portion of the NS surface around the magnetic poles, heated by a flow of relativistic particles in the pulsar magnetosphere \citep{Harding Muslimov 2002}--and non-thermal origin, generally highly beamed and sharply pulsed emission attributed to the pulsar magnetosphere, typically described by a power law with a photon index {\raise.17ex\hbox{$\scriptstyle\sim$}}1.1-1.2 \citep{Becker Trumper 1999, Zavlin 2007}, or non-pulsed emission from a shock between the pulsar wind and a flow of matter from a nondegenerate companion, as seen in ``redback'' or ``black widow'' MSPs \citep{Stappers03,Bogdanov05,Gentile14,Roberts15}. Hydrogen atmosphere models have been quite competent at describing X-ray spectra and rotation-induced pulsations of the nearby MSPs that exhibit thermal radiation \citep{Zavlin Pavlov 1998, Bogdanov Rybicki Grindlay 2007, Bogdanov Grindlay 2009}. The high density of MSPs in GCs, and their well-known distances and reddening make them ideal targets to study the relation of thermal radiation from MSPs to other pulsar parameters \citep{Kargaltsev et al. 2012}. Additionally, a more complete study of MSPs is possible from X-ray observations, as compared to other wavelengths, because X-rays from surface hot spots of these highly compact MSPs will be bent by gravity to allow observers to see {\raise.17ex\hbox{$\scriptstyle\sim$}}75\% of the neutron star surface \citep{Pechenick et al. 1983, Beloborodov 2002, Bogdanov et al. 2006}, so that a significant number of MSPs whose radio beams do not intercept the Earth should still be detectable in X-rays.

X-ray studies of MSPs also allow interesting constraints on the thermal physics of their cores. 
In addition to external return current heating, MSPs can be heated up by internal heating mechanisms, namely,  superfluid vortex creep \citep{Alpar_etal84},
rotochemical heating \citep{Reisenegger95}, 
rotation-induced deep crustal heating \citep{gkr15}, and heating produced by the dissipation of unstable oscillation modes (e.g., r-modes, 
\citealt*{ajk02,rb03,cgk17,Schwenzer_etal_Xray}).
Assuming that an MSP is in thermal balance (which is reasonable, because the thermal evolution timescale is much shorter than the spin down time-scale for MSPs),  the total power of heating should be compensated by  cooling, which depends on the MSP temperature.
Thus estimates  of MSP temperatures (even the upper limits) can be used to constrain heating processes
(see, e.g., \citealt*{Schwenzer_etal_Xray,cgk17,Mahmoodifar17}).

47 Tuc is a massive (\textit{M} {\raise.17ex\hbox{$\scriptstyle\sim$}} 10$^6$ $M_{\sun}$, \citeauthor{Pryor Meylan 1993} \citeyear{Pryor Meylan 1993}) globular cluster with a relatively high stellar concentration, although it is not core-collapsed \citep{Harris 1996}. It is considered to possess a significant population of binaries whose properties have been altered by close encounters with other stars or binaries, by virtue of being one of the clusters with the highest predicted close encounter frequencies \citep{Verbunt Hut 1987, Pooley et al. 2003}. We use a cluster absorbing column of 3.5$\times$10$^{20}$ cm$-^2$ \citep[][2010 revision]{Harris 1996}, a distance of 4.53 kpc \citep{Bogdanov et al. 2016, Hansen et al. 2013}, and metal abundances compiled in \citet{Heinke et al. 2006}. Measuring the various binary populations of clusters like 47 Tuc is crucial for understanding and modelling the dynamical encounters between binaries that produce X-ray sources in globular clusters (e.g. \citealt{Verbunt Meylan 1988}). 
Observations with the ACIS instrument of the \textit{Chandra X-Ray Observatory} in 2000 and 2002 have resulted in the identification of 300 X-ray sources within the half-mass radius of 47 Tuc \citep[hereafter H05]{Grindlay et al. 2001a, Heinke et al. 2005}. 47 Tuc is also a subject of intense scrutiny due to the large number of MSPs residing within the cluster. There have been 25 MSPs discovered so far using the 64-m Parkes Radio Telescope \citep{Manchester et al. 1990, Manchester et al. 1991, Robinson et al. 1995, Camilo et al. 2000, Pan et al. 2016, Freire et al. 2017}, with radio timing positions known for 23 of them \citep[][Freire \& Ridolfi 2017, in preparation]{Freire et al. 2001b, Freire et al. 2003, Pan et al. 2016, Ridolfi16, Freire et al. 2017}. 

Identification of optical counterparts to X-ray sources in 47 Tuc has relied on the exquisite angular resolution of the \textit{Hubble Space Telescope} (HST).  Three X-ray sources with \textit{ROSAT} positions were identified with HST counterparts \citep{Paresce et al. (1992), Paresce de Marchi 1994, Shara96}, but large numbers of identifications became possible with the subarcsecond angular resolution of \textit{Chandra}.  The deep HST program GO-8267, searching for photometric variability \citep{Gilliland00}, enabled \citet[hereafter A01]{Albrow et al. 2001}  to assemble a large catalogue of binaries exhibiting variability, largely dominated by BY Dra variables, with admixtures of short-period eclipsing variables, W UMa contact binaries, ``red stragglers'', and other variables.  BY Dra variables are pairs of main-sequence stars that have enhanced chromospheric variability compared to other stars of the same age, due to their more rapid rotation, produced by tidal locking; these stars also tend to be X-ray sources (e.g. \citealt{Dempsey97}). The combination of {\it Chandra} positions and HST data have enabled the detection of scores of optical/ultraviolet counterparts to X-ray sources, including 42 cataclysmic variables \citep{Grindlay et al. 2001a,Edmonds et al. 2003,Edmonds03b,RiveraSandoval17}, 61 chromospherically active binaries \citep{Grindlay et al. 2001a,Edmonds et al. 2003,Edmonds03b,Heinke et al. 2005,Knigge06}, 6 companions to radio millisecond pulsars \citep{Edmonds01,Edmonds02a,Edmonds et al. 2003,RiveraSandoval15,Cadelano15}, a quiescent neutron star low-mass X-ray binary \citep{Edmonds02b}, and a candidate black hole binary \citep{Bahramian et al. 2017}. 

Identifications of optical counterparts to X-ray sources typically rely on variability and/or unusual colors. Cataclysmic variables are generally bluer than the main sequence (especially in the ultraviolet), with strong H-$\alpha$ emission and variability; two of these three properties suffices to clearly identify a counterpart. Chromospherically active binaries lie up to 0.75 magnitudes above the main sequence (due to combining light from two stars), and generally show weak H-$\alpha$ emission and variability.  Millisecond pulsars and quiescent low-mass X-ray binaries show faint blue, variable counterparts, sometimes with H-$\alpha$ emission, and typically require extra information from X-ray spectra or detection of radio pulsations to distinguish from cataclysmic variables. These methods have been used to identify numerous optical/UV counterparts in 47 Tuc (references above) and in numerous other clusters \citep[e.g.][]{Grindlay01b,Pooley02b,Kong06,Bassa08,Lu11,Cohn10,Cool13}.

We combined the 2014-15 \textit{Chandra} ACIS observations \citep{Bogdanov et al. 2016} with those made in 2000 and 2002, to obtain a deeper image of 47 Tuc with improved angular resolution. In this paper, we describe the X-ray analysis we used to create a larger source catalogue with accurate source positions, and focus on the X-ray properties of the MSPs 47 Tuc F, S, Z, aa and ab, whose X-ray counterparts we have identified in this work. We have also identified new X-ray counterparts to five chromospherically active binaries previously identified by A01. We have left searches of our X-ray error circles for additional optical/UV counterparts to future works.  Finally, we used X-ray spectral fitting of the fast-spinning and X-ray dim MSP 47 Tuc aa to place tight constraints on r-mode heating processes.


\section{Observations and data reduction}
\label{sec:2}
We used data from the 2000, 2002, and 2014-15 \textit{Chandra} ACIS observations of the globular cluster 47 Tuc\footnote{We omitted the 2005-6 \textit{Chandra} observations of 47 Tuc taken with the HRC-S camera, since these have significantly higher background.}. While the five 2000 observations (described in \citealt{Grindlay et al. 2001a}) were carried out with the \textit{ACIS-I} CCD array at telescope focus, the eight 2002 observations (described in H05) and the six 2014-15 observations (described in \citealt{Bogdanov et al. 2016}) were acquired with the \textit{ACIS-S} CCD array. All the 2014-15 observations, as well as short observations in 2000 and 2002, were taken using a subarray, which reduced the frame time, and thus reduced ``pile-up''--the (incorrect) recording of two photons which landed on nearby pixels during one frame as a single event.  Subarray observations, however, only read out a portion of the CCD, encompassing the core of 47 Tuc. All the observations are summarized in Table~\ref{table:1}. The total \textit{Chandra} exposure time of 47 Tuc is 540 ks.

\begin{table}
	\caption{Summary of \textit{Chandra} observations}
	\adjustbox{max width=\columnwidth}{
	\begin{tabular}{lllll}
		\hline
		ObsID&Start Time&Exposure &Aim pt.&CCDs \\
		& & (ks) & & \\
		\hline
		78&2000-03-16 07:18:30& 3.87 & ACIS-I & 1/4\\
		953&2000-03-16 08:39:44& 31.67 & ACIS-I & 6\\
		954&2000-03-16 18:03:03& 0.85 & ACIS-I & 1/8\\
		955&2000-03-16 18:33:03& 31.67 & ACIS-I & 6\\
        956&2000-03-17 03:56:23& 4.69 & ACIS-I & 1/4\\
	    2735&2002-09-29 16:59:00& 65.24 & ACIS-S & 5\\
        3384&2002-09-30 11:38:22& 5.31 & ACIS-S & 1/4\\
        2736&2002-09-30 13:25:32& 65.24 & ACIS-S & 5\\
        3385&2002-10-01 08:13:32& 5.31 & ACIS-S & 1/4\\
        2737&2002-10-02 18:51:10& 65.24 & ACIS-S & 5\\
        3386&2002-10-03 13:38:21& 5.54 & ACIS-S & 1/4\\
        2738&2002-10-11 01:42:59& 68.77 & ACIS-S & 5\\
        3387&2002-10-11 21:23:12& 5.73 & ACIS-S & 1/4\\
        15747&2014-09-09 19:32:57& 50.04 & ACIS-S & 1/8\\
        15748&2014-10-02 06:17:00& 16.24 & ACIS-S & 1/8\\
        16527&2014-09-05 04:38:37& 40.88 & ACIS-S & 1/8\\
        16528&2015-02-02 14:23:34& 40.28 & ACIS-S & 1/8\\
        16529&2014-09-21 07:55:51& 24.7 & ACIS-S & 1/8\\
        17420&2014-09-30 22:56:03& 9.13 & ACIS-S & 1/8\\
		\hline
	\end{tabular}\label{table:1}
	}
\ \\\textit{Note}-Subarrays are indicated by fractional numbers of CCDs.
\end{table}

\begin{figure*}
	\centering
	\includegraphics[width=\textwidth]{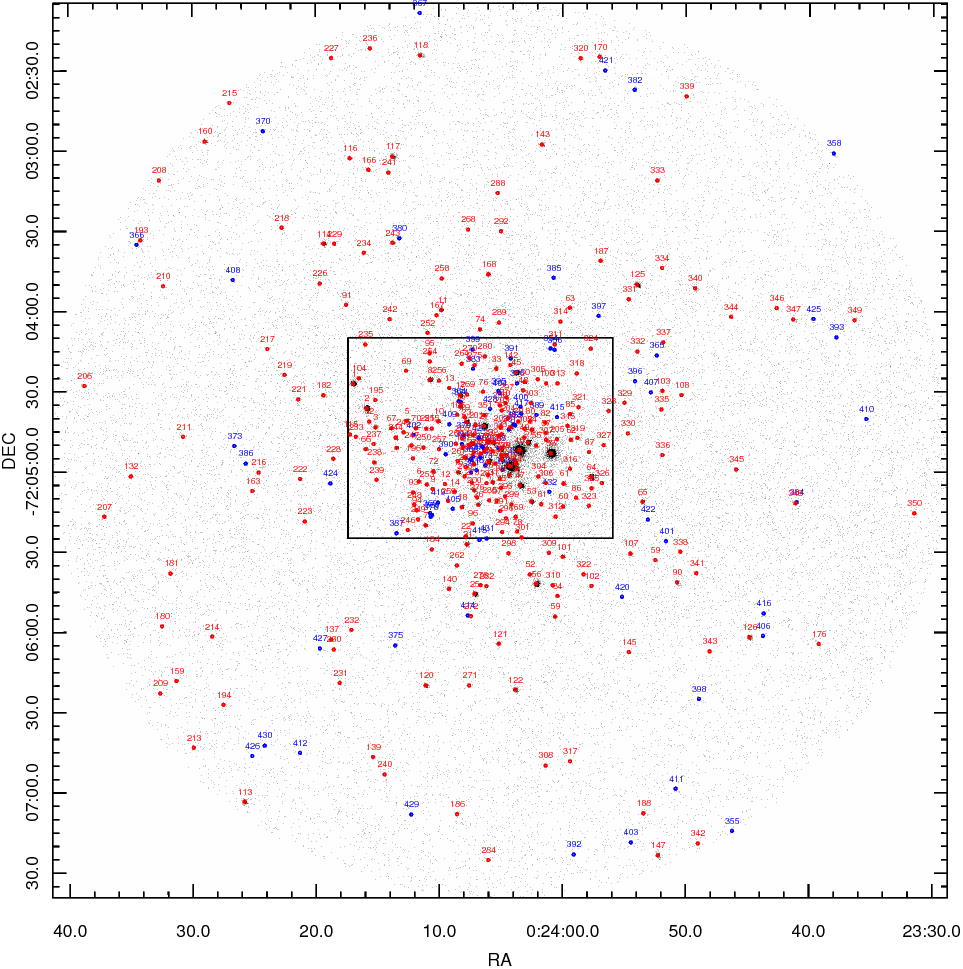}
    \caption{Combined 0.5-6 keV image of all \textit{Chandra} ACIS observations of 47 Tuc, binned to half a pixel. The centres of the circles represent the positions, while the labels indicate the W numbers of the X-ray sources. The sources shown in red were already identified by H05 while those in blue have been identified in this work. The sources within the inset box are shown in Fig.~\ref{fig:2}.}
    \label{fig:1}
\end{figure*}

\begin{figure*}
    \centering
	\includegraphics[width=\textwidth]{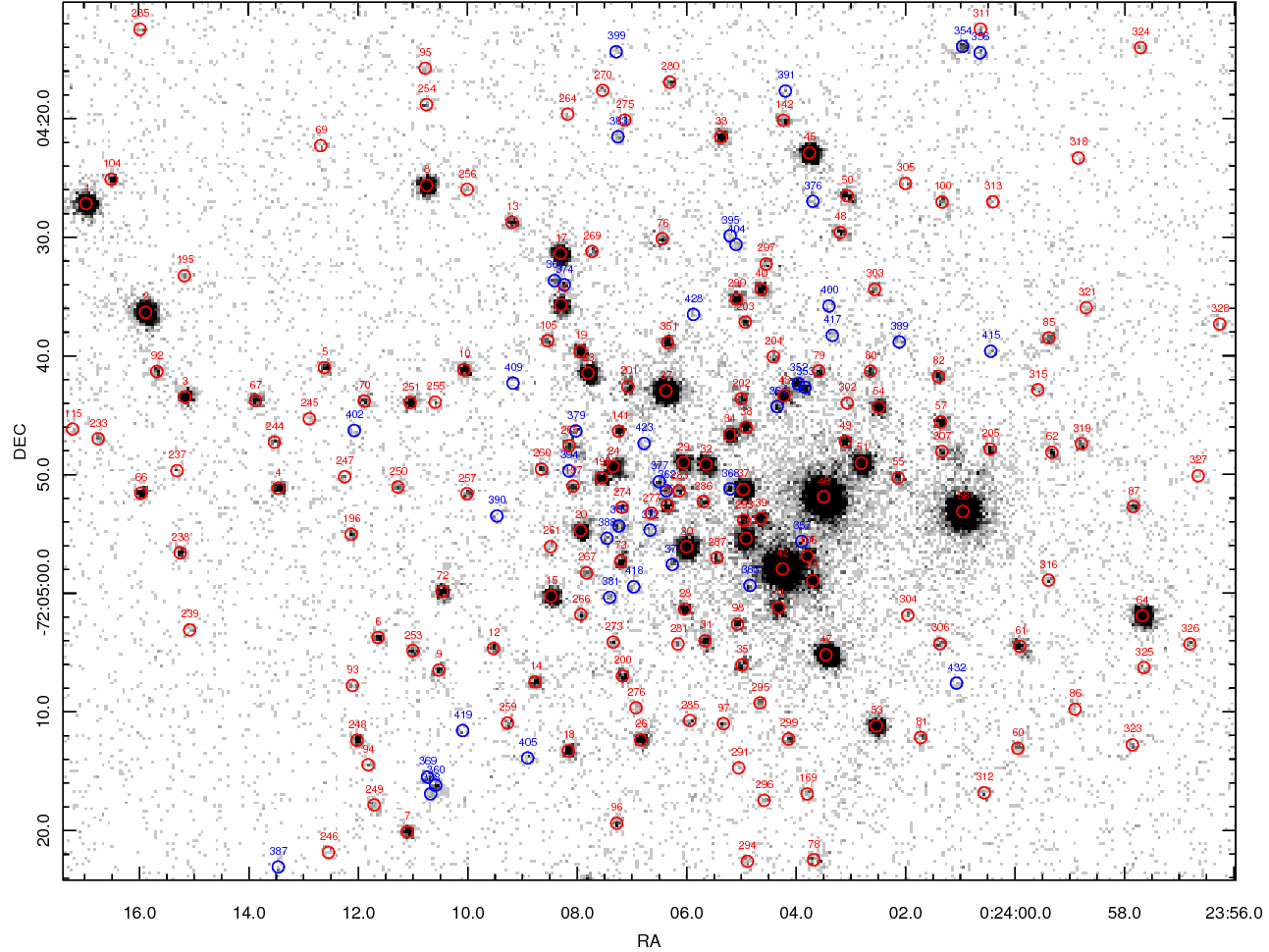}
    \caption{Same as that of Fig.~\ref{fig:1} but within the inset box.}
    \label{fig:2}
\end{figure*}

The data were reduced using CIAO version 4.8\footnote{http://cxc.cfa.harvard.edu/ciao/} and CALDB version 4.7.1, in accordance with standard CIAO science threads\footnote{http://cxc.harvard.edu/ciao/threads/index.html}. All the observations were reprocessed from the original level 2 event files following the default ACIS reprocessing steps. The reprocessing applies the sub-pixel event-repositioning algorithm EDSER\footnote{http://cxc.harvard.edu/ciao/why/acissubpix.html} \citep{Li et al. 2004}, which improves on-axis image resolution, thereby improving the resolution for the crowded centre of 47 Tuc.  {The intrinsic on-axis PSF of the Chandra mirrors for a typical spectrum is reasonably represented by a Gaussian with FWHM of $0.4''$ (Chandra POG\footnote{http://cxc.cfa.harvard.edu/proposer/POG/html/}, chapter 7). The ACIS detector pixels are $0.492''$ in size, so oversampling them is clearly beneficial. The effects of using the EDSER algorithm on the Chandra PSF are yet to be calibrated.} We limited the energy range to 0.5-6 keV and covered the cluster out to 2.79 arcminutes, the half-mass radius used by Heinke et al. 2005, in each observation. (We use this limiting radius for compatibility with Heinke et al. 2005 and other Chandra studies of globular clusters, motivated by the dominance of non-cluster sources outside that radius.) We matched the astrometry of all the observations to ObsID 2735.  {Pile-up at the level of 10-15 \% (the maximum seen for any source in 47 Tuc) has a very small effect on astrometric corrections, as proven by the lack of detectable differences in alignment with their optical counterparts between the brightest X-ray sources and fainter X-ray sources in 47 Tuc \citep{Edmonds et al. 2003}.}
We merged all the observations to obtain a deeper event file and subsequently produced images binned to a quarter of an arcsecond (half a pixel), for source detection. We also created exposure maps and aspect histograms for all the CCDs, for each observation, to use with the ACIS-EXTRACT (AE) package\footnote{http://www2.astro.psu.edu/xray/docs/TARA/ae\textunderscore users\textunderscore guide.html}, version 2016feb1, as discussed later.

\section{Analysis}
\subsection{Source Detection}
\label{sec:2.2}
We employed two source detection algorithms, CIAO's \textit{wavdetect}\footnote{http://cxc.harvard.edu/ciao/threads/wavdetect/} algorithm \citep{Freeman et al. 2002}, and the independent \textit{pwdetect}\footnote{http://www.astropa.unipa.it/progetti\textunderscore ricerca/PWDetect/} algorithm \citep{Damiani et al. 1997}. The \textit{wavdetect} tool employs a Mexican-Hat wavelet-based source detection algorithm that detects probable sources within a dataset using significant correlations of source pixels with wavelets of different scales. For this, we created images binned to half a pixel in the 0.5-6 keV, 0.5-2 keV and 2-6 keV energy bands, using scales of 1.414, 2.0, 2.828, 4, 5.656 and 8.0 pixels, with a source detection significance threshold of 10$^{-6}$, which should result in one false detection per ACIS chip. We chose to use larger source detection scales in order to permit the detection of point sources further away from the aimpoint but within the half-mass radius. The half-pixel binning of the images ensured that even in the crowded centre of 47 Tuc, \textit{wavdetect} could separate close sources in spite of the larger source detection scales. 

\textit{pwdetect} is also a wavelet-based source detection algorithm which performs a multiscale analysis of the data, thus allowing the detection of both pointlike and moderately extended sources in the entire field of view.  {Compared to \textit{wavdetect}, it is more effective in the detection of faint sources close to brighter ones (as seen by \citealt{Heinke et al. 2003a} and \citealt{Forestell et al. 2014})}. To use \textit{pwdetect}, we created images binned to one pixel in the aforementioned energy bands, using scales from 0.5$''$ to 1$''$, and a final detection threshold of 5.1$\sigma$, which should also result in one false detection per ACIS chip. We also created a merged event file consisting exclusively of split-pixel events, as this is known \citep[e.g.][]{Zurek et al. 2016} to improve Chandra's angular resolution at the cost of {\raise.17ex\hbox{$\scriptstyle\sim$}}25\% of the total counts. Although, in our case, we lost about {\raise.17ex\hbox{$\scriptstyle\sim$}}33\% of the total counts.

We found an improved performance for \textit{wavdetect} as compared to previous reports (e.g. H05) in separating sources in the crowded cluster centre, which we attribute to the EDSER algorithm and use of the half-pixel binned image. We also corroborated the performance of \textit{pwdetect} in detecting faint sources close to bright ones. Despite the 12 detection runs, an additional 7 possible sources near the crowded cluster centre can still be clearly identified by eye. We created a combined source list with all the sources identified from the detection runs and by eye. Subsequent comparison with the source list from H05 leads us to identify 20 additional sources. We add these to our source catalogue as well, and refine our combined source catalogue further using the AE package, detailed by \citet{Broos et al. 2010}. 

Spectra and background were extracted for each source in the catalog using the AE package (explained in detail in \S \ref{sec:2.3}). Subsequently, the positions of these sources were refined by calculating the centroid of the data within a preliminary extraction region. When the probability of the extracted counts being produced by fluctuations in the background (\textit{PROB\textunderscore NO\textunderscore SOURCE}) was above a threshold value of {\raise.17ex\hbox{$\scriptstyle\sim$}}10\% \citep{Weisskopf et al. 2007}, the source was removed from the catalogue, followed by a refinement of the positions. This process was repeated till no further pruning of the catalogue was necessary. 

\begin{landscape}
\begin{table}
	\caption{47 Tuc basic X-ray source properties, combined data-set}
	\adjustbox{width=\columnwidth}{
	\begin{tabular}{ccccccccccc}
		\hline
		\multicolumn{2}{c}{NAME} & \multicolumn{3}{c}{POSITION}& & & \multicolumn{2}{c}{COUNTS}& LUMINOSITY& \\
		\hline
		Label&CXOGlb J&$\alpha$ &$\delta$&P$_{\rm err}$&f$_{\rm PSF}$&f$_{\rm EXP}$&0.5-2 keV&0.5-6 keV&0.5-6 keV&Notes \\
		 & &(h:m:s) &(${\circ}:':''$) &(arcsec) & & & & & (10$^{30}$ ergs s$^{-1}$) & \\
		\hline
42&002404.3-720458& 00:24:04.251&-72:04:57.976&0.277&0.90&1.00&20696.0$^{+205.7
}_{-203.6}$ &29194.4$^{+335.8}_{-331.7}$ &1151.9$^{+13.2}_{-13.1}$ &X9\\
46&002403.5-720452& 00:24:03.500&-72:04:51.888&0.276&0.90&1.00&43143.4$^{+296.4
}_{-294.3}$ &45417.4$^{+352.1}_{-294.3}$ &1027.9$^{+8.0}_{-6.7}$ &X7\\
58&002401.0-720453& 00:24:00.956&-72:04:53.137&0.278&0.90&1.00&23852.4$^{+220.6
}_{-218.5}$ &25271.7$^{+267.3}_{-255.5}$ &574.2$^{+6.1}_{-5.8}$ &X5\\
56&002402.1-720542& 00:24:02.125&-72:05:41.982&0.286&0.90&0.98&2428.3$^{+70.7
}_{-68.6}$ &3392.6$^{+115.2}_{-111.1}$ &169.5$^{+5.8}_{-5.6}$ &X6\\
27&002406.4-720443& 00:24:06.378&-72:04:42.956&0.284&0.90&1.00&4729.0$^{+98.5
}_{-96.4}$ &5555.4$^{+140.2}_{-136.1}$ &167.7$^{+4.2}_{-4.1}$ &X10\\
47&002403.5-72055& 00:24:03.456&-72:05:05.212&0.287&0.90&1.00&2167.9$^{+66.9}_{-
64.8}$ &3074.0$^{+109.6}_{-105.5}$ &113.4$^{+4.0}_{-3.9}$ &-\\
2&002415.9-720436& 00:24:15.880&-72:04:36.336&0.289&0.90&0.98&1529.5$^{+56.0}_{-
54.0}$ &2250.7$^{+94.6}_{-90.5}$ &96.1$^{+4.0}_{-3.9}$ &X13\\
1&002417.0-720427& 00:24:16.966&-72:04:27.160&0.290&0.90&0.97&1478.9$^{+55.9}_{-
53.9}$ &1970.5$^{+88.1}_{-83.9}$ &75.3$^{+3.4}_{-3.2}$ &-\\
36&002404.9-720455& 00:24:04.915&-72:04:55.402&0.295&0.90&1.00&235.1$^{+26.8}_{-
24.3}$ &977.5$^{+67.5}_{-62.8}$ &73.0$^{+5.0}_{-4.7}$ &-\\
30&002406.0-720456& 00:24:06.001&-72:04:56.104&0.290&0.90&1.00&1456.7$^{+55.6
}_{-53.5}$ &1983.3$^{+89.0}_{-84.8}$ &71.6$^{+3.2}_{-3.1}$ &X19\\
45&002403.8-720423& 00:24:03.759&-72:04:22.912&0.291&0.90&0.87&1302.6$^{+52.6
}_{-50.6}$ &1700.8$^{+81.8}_{-77.7}$ &64.6$^{+3.1}_{-3.0}$ &-\\
64&002357.7-72052& 00:23:57.678&-72:05:01.918&0.292&0.90&0.99&964.4$^{+45.6}_{-
43.6}$ &1400.9$^{+76.8}_{-72.7}$ &58.1$^{+3.2}_{-3.0}$ &-\\
51&002402.8-720449& 00:24:02.810&-72:04:49.066&0.291&0.90&1.00&1232.8$^{+52.3
}_{-50.1}$ &1629.5$^{+81.8}_{-77.4}$ &56.7$^{+2.8}_{-2.7}$ &-\\
25&002407.1-720546& 00:24:07.139&-72:05:45.730&0.294&0.90&0.98&854.5$^{+43.2}_{-
41.1}$ &1112.6$^{+67.2}_{-63.0}$ &55.6$^{+3.4}_{-3.2}$ &X11\\
8&002410.7-720426& 00:24:10.747&-72:04:25.644&0.295&0.89&0.96&380.0$^{+27.7}_{-
25.6}$ &904.8$^{+61.6}_{-57.5}$ &54.4$^{+3.7}_{-3.5}$ &-\\
23&002407.8-720441& 00:24:07.802&-72:04:41.473&0.292&0.89&1.00&1060.0$^{+46.9
}_{-44.8}$ &1376.8$^{+72.0}_{-67.9}$ &43.8$^{+2.3}_{-2.2}$ &-\\
125&002354.0-720350& 00:23:53.985&-72:03:50.069&0.294&0.90&0.82&951.5$^{+45.3
}_{-43.2}$ &1038.9$^{+60.2}_{-55.9}$ &43.1$^{+2.5}_{-2.3}$ &X4\\
15&002408.5-72050& 00:24:08.477&-72:05:00.269&0.297&0.90&0.98&330.3$^{+25.3}_{-
23.2}$ &711.0$^{+54.4}_{-50.2}$ &39.4$^{+3.0}_{-2.8}$ &-\\
53&002402.5-720511& 00:24:02.533&-72:05:11.202&0.296&0.90&1.00&605.2$^{+36.6}_{-
34.5}$ &846.7$^{+60.2}_{-56.0}$ &32.0$^{+2.3}_{-2.1}$ &-\\
37&002405.0-720451& 00:24:04.966&-72:04:51.272&0.292&0.89&1.00&1290.6$^{+53.5
}_{-51.4}$ &1360.9$^{+67.2}_{-59.1}$ &31.9$^{+1.6}_{-1.4}$ &-\\
32&002405.6-720449& 00:24:05.647&-72:04:49.159&0.295&0.90&1.00&768.2$^{+41.6}_{-
39.5}$ &937.7$^{+61.7}_{-57.4}$ &29.5$^{+1.9}_{-1.8}$ &-\\
17&002408.3-720431& 00:24:08.306&-72:04:31.393&0.294&0.90&0.99&916.0$^{+44.8}_{-
42.7}$ &1042.5$^{+62.1}_{-57.8}$ &28.8$^{+1.7}_{-1.6}$ &-\\
114&002419.4-720335& 00:24:19.364&-72:03:34.729&0.301&0.90&0.90&397.0$^{+30.2
}_{-28.2}$ &492.2$^{+45.8}_{-41.5}$ &24.5$^{+2.3}_{-2.1}$ &-\\
29&002406.1-720449& 00:24:06.059&-72:04:49.004&0.301&0.90&1.00&357.3$^{+29.2}_{-
27.0}$ &495.2$^{+47.8}_{-43.6}$ &19.7$^{+1.9}_{-1.7}$ &MSP-W\\
117&002413.8-72032& 00:24:13.794&-72:03:02.178&0.304&0.90&0.96&293.4$^{+26.2}_{-
24.1}$ &372.6$^{+40.7}_{-36.4}$ &18.5$^{+2.0}_{-1.8}$ &-\\
122&002403.8-720622& 00:24:03.847&-72:06:21.586&0.304&0.90&0.99&286.0$^{+26.0
}_{-23.9}$ &363.1$^{+40.1}_{-35.8}$ &18.3$^{+2.0}_{-1.8}$ &-\\
16&002408.3-720436& 00:24:08.298&-72:04:35.720&0.302&0.90&0.99&310.2$^{+26.9}_{-
24.8}$ &449.9$^{+45.4}_{-41.2}$ &18.1$^{+1.8}_{-1.7}$ &-\\
120&002411.1-720620& 00:24:11.099&-72:06:19.973&0.307&0.90&0.96&214.0$^{+22.8
}_{-20.7}$ &287.2$^{+36.9}_{-32.6}$ &17.1$^{+2.2}_{-1.9}$ &-\\
44&002403.7-720459& 00:24:03.694&-72:04:58.980&0.303&0.90&1.00&297.3$^{+27.4}_{-
25.2}$ &414.4$^{+44.7}_{-40.3}$ &15.5$^{+1.7}_{-1.5}$ &-\\
126&002344.8-72062& 00:23:44.828&-72:06:01.976&0.312&0.90&0.81&107.7$^{+15.6}_{-
13.4}$ &208.8$^{+31.5}_{-27.1}$ &15.5$^{+2.3}_{-2.0}$ &-\\
24&002407.3-720449& 00:24:07.339&-72:04:49.325&0.300&0.89&1.00&463.5$^{+32.5}_{-
30.5}$ &532.9$^{+46.0}_{-41.7}$ &14.8$^{+1.3}_{-1.2}$ &-\\
20&002407.9-720455& 00:24:07.938&-72:04:54.736&0.300&0.90&1.00&429.7$^{+31.0}_{-
28.9}$ &515.0$^{+45.2}_{-40.9}$ &14.7$^{+1.3}_{-1.2}$ &-\\
		\hline
	\end{tabular}\label{table:2}
    }
\ \\\textit{Note}-The source positions have been adjusted to place them on to the radio frame, using the X-ray detections of 19 radio MSPs. Positional errors are quoted in arcseconds for both RA and DEC, and are the 95\% error circles as calculated by \citet{Hong et al. 2005}. The PSF fraction (f$_{\rm PSF}$) and the fractional exposure time (f$_{\rm EXP}$) have also been mentioned for each source. The Notes column indicates \textit{c} for sources which appear confused, and \textit{m} for sources added manually, and includes other reported names (X for ROSAT sources, MSP-* for MSPs). A portion of this table is shown here for guidance, the full table is available in the electronic edition of the Journal.
\end{table}
\end{landscape}

This pruning left us with 370 X-ray sources detected within the cluster half-mass radius, as catalogued in Table~\ref{table:2}  { (a portion of this table is shown here for guidance, the full table is available in the electronic edition of the Journal). The source positions are centroid positions which were further corrected by matching the astrometry with that of the previously known MSPs in 47 Tuc (detailed in \S \ref{sec:2.4}). For the positional error of each source, we quote the radius of the 95\% error circle (P$_{\rm err}$), calculated using the empirical formula derived by \citet{Hong et al. 2005} from applying \textit{wavdetect} on simulated data.} Here the source positions were ordered by decreasing luminosity in the 0.5-6 keV band, and have been labelled accordingly. We retained the W numbering scheme of \citet{Grindlay et al. 2001a}, as extended in H05 to cover all the sources previously identified from the 2000 and 2002 observations, extending it further to cover the additional ones identified in this work. The W numbers for 5 sources which have been resolved into multiple sources, and 6 sources which were previously identified but were not detected in our analysis, have been omitted. 
Sources with \textit{PROB\textunderscore NO\textunderscore SOURCE} greater than {\raise.17ex\hbox{$\scriptstyle\sim$}}1.5\% but less than {\raise.17ex\hbox{$\scriptstyle\sim$}}10\% have been retained in our final catalogue, but since their detection is marginal, they have been marked as \textit{c} in Table~\ref{table:2}. The sources identified by eye have been marked as \textit{m}.

\begin{figure}
    \centering
	\includegraphics[width=\columnwidth]{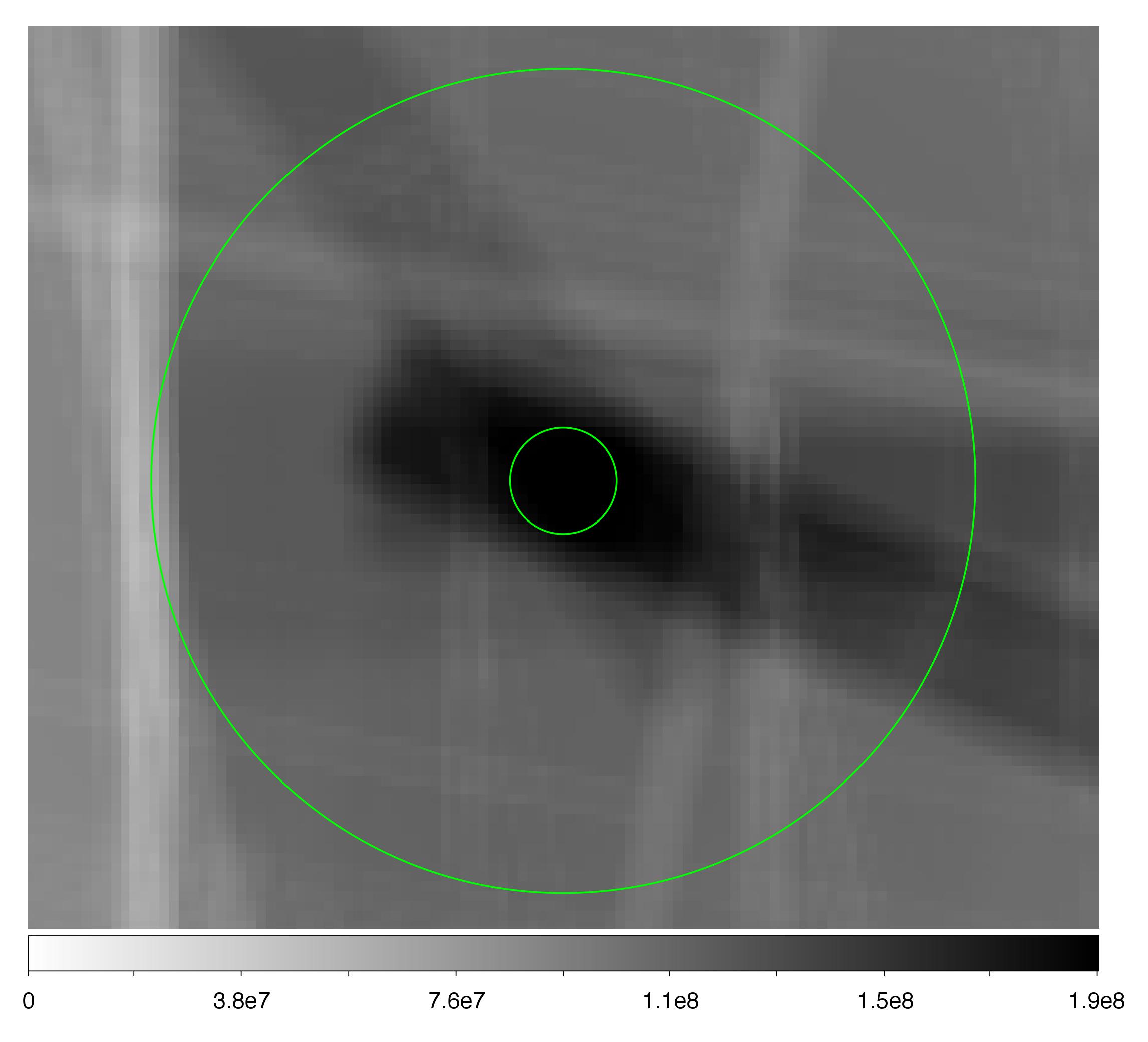}
    \caption{Combined exposure map from the 19 merged observations of 47 Tuc  {in units of cm$^2$s}. The core (0.36') and half-mass (2.79', as used by H05) radii are marked for reference. Features in the exposure map include reduced exposure due to a chip gap at the east edge of the half-mass radius from the 2002 observations, chip gaps crossing the interior from the 2000 observations, and regions of increased exposure due to the positions of the 2014-2015 subarray observations (all of which covered the cluster core).}
    \label{fig:2.5}
\end{figure}

\begin{figure*}
    \centering
	\includegraphics[width=\textwidth]{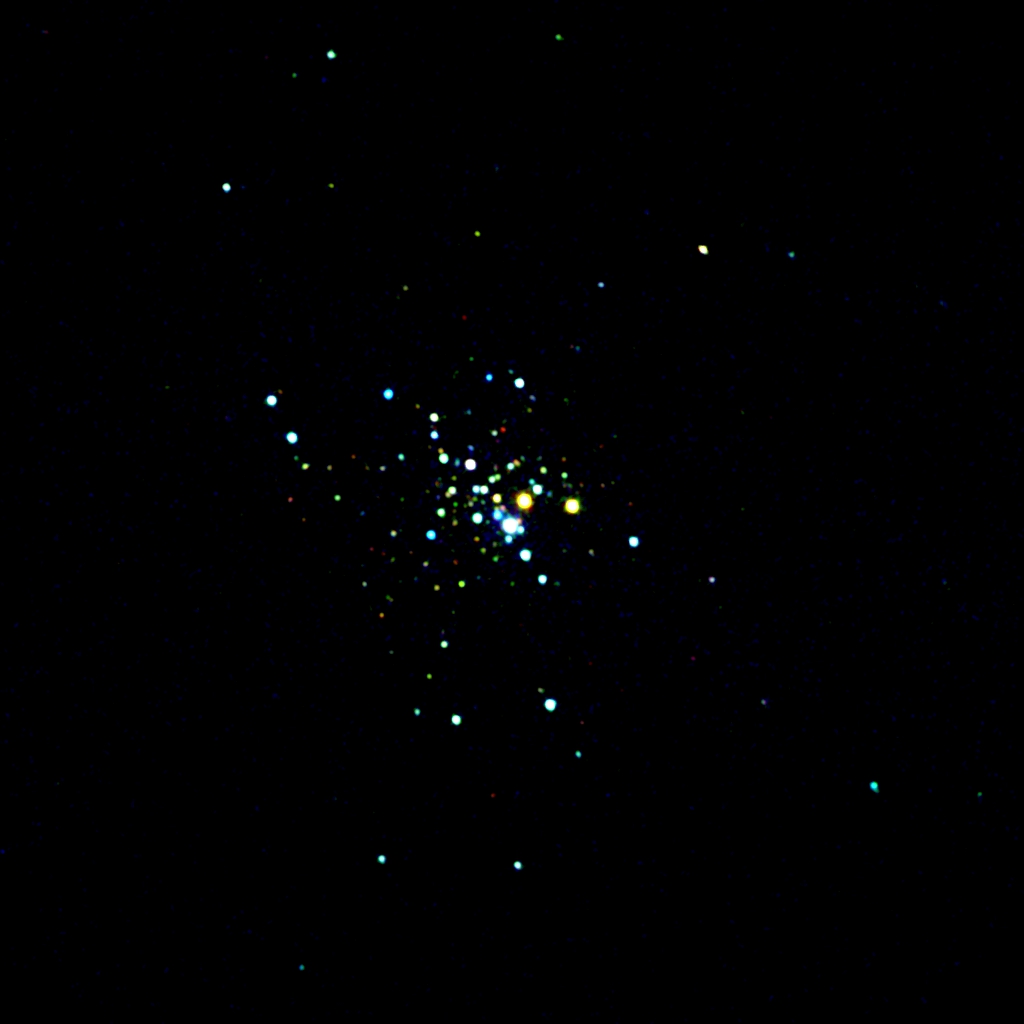}
    \caption{True colour image for the 19 merged observations of 47 Tuc within the half-mass radius. It was constructed from a 0.2-1.5 keV image (red), a 1.5-2.5 keV image (green), and a 2.5-8 keV image (blue). All images were binned 
to 0.25\arcsec\ pixels, smoothed using ds9 with a gaussian kernel of radius 3 pixels, and then combined.}
    \label{fig:3}
\end{figure*}

 {Fig.~\ref{fig:1} identifies all the sources found within the cluster half-mass radius, while Fig.~\ref{fig:2} identifies those within the crowded core, both plotted on an image produced by combining data from the 19 ACIS observations binned to half a pixel in the 0.5-6 keV energy band.} The sources shown in red were already identified by H05 while those in blue, numbering 81, have been newly identified in this work.  {The newly identified sources are either fainter than the previously identified ones, or are located close to bright sources.} Fig.~\ref{fig:2.5} shows a merged exposure map (in units of cm$^2$ s, encoding the telescope effective area and the amount of time each location was imaged) covering the half-mass radius, illustrating that we obtain the highest sensitivity in the core. 
Fig.~\ref{fig:3} shows a representative true-colour image of the combined data obtained from the 19 merged observations made from the 0.2-1.5, 1.5-2.5, and 2.5-8 keV data.

\subsection{Extraction and photometry}
\label{sec:2.3}
From the positions defined in our initial source catalogue, source and background spectra were extracted. Events were selected from each observation, for each source, from within a region that encompassed 90 percent of the Point Spread Function (PSF) centred on each catalogue position, or a region of reduced size if the sources were too crowded.  {The typical on-axis extraction radius was $\sim 0.9''$}.
 From these events, source and background spectra were extracted by the ae\textunderscore standard\textunderscore extraction tool, which also generated effective area files (ARFs) and response matrices (RMFs) by calling the appropriate CIAO tools. Background extractions included at least 50 counts, involving masks designed to accurately assess the local background due to neighbouring point sources as well as the instrumental background, and sampled pixels from areas outside all source extraction regions \citep[for details, see][]{Broos et al. 2010}. 

After extracting photons, AE runs the CHECK\textunderscore POSITIONS stage, which can produce estimates of the source position by several methods, including performing image reconstruction on a crowded field with a maximum likelihood method \citep{Broos et al. 2010}. We obtained these image reconstruction positions for faint sources that suffered significant crowding in the core region; these objects have been listed in  Table~\ref{table:3}.  {The average shift between these positions and the centroid positions quoted in Table~\ref{table:2} for the corresponding sources are 0.34$''$ in RA and 0.13$''$ in DEC.}

These extractions were repeated until the positions of sources were well-determined and no further pruning was needed. Background-subtracted photometry was calculated in several bands. We also determined the number of counts for each catalogue source in the 0.5-2 and 2-6 keV bands, and also photon fluxes (quoted as ``FLUX2'' in AE; that is, net counts divided by the mean effective area in the band, and by the exposure time) in the 0.5-1, 1-2, 2-4 and 4-6 keV bands.

\begin{table}
	\caption{Alternative positions of crowded sources, obtained from image reconstruction.}
	\centering
    \adjustbox{max width=\columnwidth}{
	\begin{tabular}{ccc}
		\hline
		Label&$\alpha$ &$\delta$ \\
		 &(h:m:s) &(${\circ}:':''$)  \\
		\hline
24	&	00:24:07.342	&	-72:04:49.303	\\
41	&	00:24:04.321	&	-72:05:01.280	\\
43	&	00:24:04.218	&	-72:04:43.403	\\
44	&	00:24:03.687	&	-72:04:58.948	\\
75	&	00:24:06.331	&	-72:04:52.626	\\
106	&	00:24:03.786	&	-72:04:56.870	\\
198	&	00:24:07.577	&	-72:04:50.308	\\
265	&	00:24:08.176	&	-72:04:47.834	\\
283	&	00:24:06.141	&	-72:04:51.384	\\
293	&	00:24:04.955	&	-72:04:53.839	\\
348	&	00:23:41.132	&	-72:05:11.252	\\
352	&	00:24:03.972	&	-72:04:42.323	\\
353	&	00:24:03.836	&	-72:04:42.708	\\
357	&	00:24:03.947	&	-72:04:55.776	\\
360	&	00:24:10.578	&	-72:05:16.260	\\
361	&	00:24:04.344	&	-72:04:44.360	\\
362	&	00:24:06.373	&	-72:04:51.384	\\
363	&	00:24:04.880	&	-72:04:59.552	\\
364	&	00:24:08.409	&	-72:04:33.704	\\
368	&	00:24:05.239	&	-72:04:51.280	\\
369	&	00:24:10.718	&	-72:05:15.608	\\
371	&	00:24:06.240	&	-72:04:57.428	\\
374	&	00:24:08.246	&	-72:04:34.104	\\
377	&	00:24:06.507	&	-72:04:50.675	\\
378	&	00:24:10.687	&	-72:05:17.117	\\
379	&	00:24:08.003	&	-72:04:46.268	\\
384	&	00:23:40.935	&	-72:05:11.242	\\
404	&	00:24:05.081	&	-72:04:30.677	\\
\hline
	\end{tabular}\label{table:3}
	}
\ \\
\end{table}

We calculated the total luminosity in the 0.5-6 keV band, using XSPEC version 12.9\footnote{http://heasarc.gsfc.nasa.gov/docs/xanadu/xspec/}. We calculated the conversion from observed photon fluxes to (unabsorbed) energy fluxes using the XSPEC VMEKAL model, with a typical temperature of 2 keV, and accounting for Galactic absorption with the TBABS model \citep{Wilms 2000}, to the combined spectrum. We chose the VMEKAL model since we expect these faint sources to be dominated by chromospherically active binaries and CVs, both of which have X-ray spectra well represented by MEKAL models (e.g. H05). Using the known cluster absorbing column, distance and metal abundances, we computed a photon cm$^{-2}$ s$^{-1}$ to erg cm$^{-2}$ s$^{-1}$ conversion for the different flux bands, and subsequently used these conversions to calculate luminosity in the 0.5-6 keV band, for all the sources (as shown in Table~\ref{table:2}).

\subsection{Astrometric corrections}
\label{sec:2.4}
 {The final refined positions of our sources were further corrected to the known radio-timing positions of 19 out of the 23 MSPs with well-constrained positions in 47 Tuc.} The known positions for 16 of these MSPs were obtained from \citet{Freire et al. 2003} while that for the MSP 47 Tuc ab was obtained from \citet{Pan et al. 2016}, for 47 Tuc W from \citet{Ridolfi16}, and for 47 Tuc Y from \citet{Freire et al. 2017}. The astrometric corrections were carried out using CIAO with a matching radius of 1$''$. 

Disregarding the counterparts of 47 Tuc S and 47 Tuc F for reasons explained in \S \ref{sec:3}, and 47 Tuc L, whose counterpart is affected by being too close to a bright source, we have constrained the positions of our sources to a standard deviation of {\raise.17ex\hbox{$\scriptstyle\sim$}}0.16$''$ in RA and {\raise.17ex\hbox{$\scriptstyle\sim$}}0.08$''$ in DEC, of the other MSPs. The deviations for individual MSPs are listed in Table~\ref{table:astro}.  {This table also includes the positions for MSPs 47 Tuc R and Z, whose positions have been reported in \citet{Freire et al. 2017}, and 47 Tuc aa, whose position will be reported in Freire \& Ridolfi (2017, in preparation). Only three of the 22 X-ray counterparts lie farther than the \citet{Hong et al. 2005} 95\% error circle from the radio position, suggesting that most of the X-ray counterparts are correct counterparts. Two of the three counterparts which are farther than expected from their radio positions (R and L) lie in the wings (1.5'' and 2.6'' distance, respectively) of brighter sources, which is likely to affect their estimated X-ray positions. }

 {We shifted the radio positions of all the MSPs by $5''$ in four directions, and searched for X-ray sources within $0.5"$ of these shifted positions, thus estimating the probability of chance superposition with unrelated sources. 
This false match probability per source per trial was found to be $0.0375$, indicating that of order one MSP is expected to coincide with another X-ray source within 47 Tuc.  This is consistent with the calculation in H05 of a 3.5\% chance for any given source above 20 counts to fall within 0.5$"$ of another such source. Any MSPs where this situation occurs would be detected as the sum of the X-ray emission of the MSP and the other source. H05 speculates that this may explain the detection of X-ray variability from 47 Tuc O, as the X-ray emission from 47 Tuc O is not expected to be variable on these timescales, while it would not be surprising, for instance, from a chromospherically active binary. }

\begin{table}
	\caption{Deviations in X-ray positions of individual MSPs from their known radio positions. }
    \adjustbox{max width=\columnwidth}{
	\begin{tabular}{llllll}
		\hline
		Label&MSP&$\Delta\alpha$ &$\Delta\delta$ & X-ray err & $\Delta$/Xerr \\
		 &&($''$) &($''$) & ($''$)  \\
		\hline
7	&   E   &	$-0.22$	&	$-0.02$ & 0.32 & 0.69	\\
11	&   U   &	$-0.06$	&	$-0.06$ & 0.33 & 0.26	\\
13	&   N   &	$-0.04$	&	$-0.07$ & 0.33 & 0.24	\\
19	&   G,I  & $0.05$, $0.2$ & $-0.02$, $-0.04$ & 0.31 & $<$0.21 \\
28	&   Z   &	$-0.01$	&	$-0.10$ & 0.32 & 0.31	\\
29	&   W   &	$0.01$	&	$-0.09$ & 0.30 & 0.30	\\
39	&   O   &	$0.12$	&	$0.00$ &	0.31 & 0.39  \\
63	&   J   &	$0.12$	&	$-0.02$ & 0.32 & 0.38	\\
67	&   D   &	$-0.08$	&	$-0.08$ & 0.32 & 0.35	\\
74	&   H   &	$0.08$	&	$0.10$ &	0.33 & 0.40 \\
82	&   Y   &	$0.00$	&	$-0.08$ & 0.33 & 0.24	\\
96	&   aa   &	$0.02$	&	$-0.12$ & 0.36 & 0.34	\\
104	&   Q   &	$0.34$	&	$-0.04$ & 0.33 & 1.04	\\
105	&   T   &	$-0.01$	&	$-0.16$ & 0.34 & 0.47	\\
106	&   L   &	$0.63$	&	$0.03$ &	0.31 & 2.03 \\
107	&   M   &	$0.08$	&	$-0.14$ & 0.34 & 0.47	\\
108	&   C   &	$0.27$	&	$-0.12$ & 0.35 & 0.84	\\
198	&   R   &	$-0.44$	&	$-0.11$ & 0.32 & 1.42	\\
265	&   ab   &	$-0.32$	&	$-0.03$ & 0.33 & 0.97	\\
352	&   S   &	$-0.25$	&	$0.03$ &	0.32 & 0.79  \\
353	&   F   &	$-0.27$	&	$-0.15$ & 0.34 & 0.64	\\

\hline
	\end{tabular}\label{table:astro}
	}
\ \\\textit{Note}- The columns give the wavdetect number, MSP radio detection name (e.g. 47 Tuc E), deviations in RA and Dec, the 95\% confidence error circle radius on the X-ray position (following \citealt{Hong et al. 2005}), and the total deviation divided by the X-ray error.
\end{table}

\subsection{Optical Counterparts}
\label{sec:3.5oc}
We identified candidate optical counterparts from the lists of variable stars given by A01 for 5 of the newly identified X-ray sources (Table~\ref{table:A01}). Astrometric corrections to the HST positions of the A01 binaries (0.269$''$ in RA and 0.085$''$ in DEC) were made by aligning HST counterparts to the five brightest X-ray sources with identifications by \citet{Edmonds et al. 2003} to their X-ray positions derived in this work. These X-ray sources were found to have a rms offset of {\raise.17ex\hbox{$\scriptstyle\sim$}}0.025$''$ in RA and {\raise.17ex\hbox{$\scriptstyle\sim$}}0.035$''$ in DEC with their corresponding astrometrically corrected HST positions. We identified variable stars from A01 as candidate optical counterparts to our newly identified X-ray sources within thrice the rms offsets in RA and DEC from the X-ray positions (0.13\arcsec; Table~\ref{table:A01}).
All the candidate optical counterparts identified in this work, except WF2-V19, were suggested by H05 to be faint and/or confused X-ray sources, which we were able to identify with our deeper X-ray source list.

\begin{table}
	\caption{New Optical Counterpart Identifications}
    \adjustbox{max width=\columnwidth}{
	\begin{tabular}{llllll}
		\hline
		W&A01 Name&$\Delta\alpha$ &$\Delta\delta$ &Type&Period \\
		 & &($''$) &($''$) & &(days) \\
		\hline
	    363&PC1-V28&$-0.036$&$0.089$&BY Dra&1.16 \\
	    368&PC1-V07&$-0.012$&$-0.027$&W UMa S.Det.&0.4188 \\
	    399&WF2-V19&$-0.038$&$0.057$&BY Dra&1.56 \\
	    414&WF4-V09&$0.005$&$0.063$&BY Dra&1.37 \\
	    424&WF3-V21&$-0.036$&$-0.049$&BY Dra&4.03 \\
		\hline
	\end{tabular}\label{table:A01}
	}
\ \\\textit{Note}-Separations ($\Delta\alpha$ \& $\Delta\delta$) are defined as Chandra minus HST positions. Types and periods are from A01.
\end{table}

\begin{figure}
    \centering
	\includegraphics[width=\columnwidth]{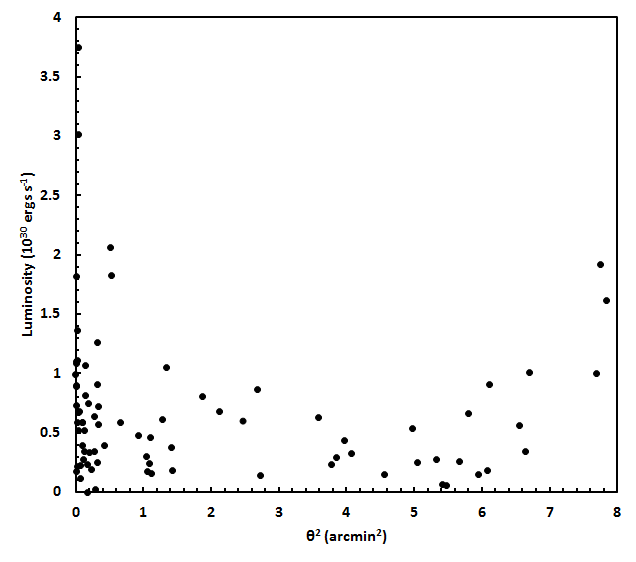}
    \caption{Distribution of luminosity vs. radial distance from the centre of 47 Tuc, plotted in units of arcmin$^2$, for sources identified by this work.  Outside $\theta^2=2$arcmin$^2$, the flat distribution suggests a uniform spatial distribution.}
    \label{fig:4}
\end{figure}

\subsection{Radial Distribution}
\label{sec:2.5}
In Fig.~\ref{fig:4}, we plotted the radial distance from the centre of 47 Tuc, $\alpha$= $00^{h}24^{m}05^{s}.29$, $\delta$= $-72^{\circ}04'52''.3$ \citep{de Marchi et al. 1996}, in units of arcmin$^2$, for the 81 sources newly identified by this work.  The sources in the core are most likely cluster members, while sources near the outskirts are probably dominated by background extragalactic sources.  The distribution of expected background extragalactic objects should be homogeneous over the half-mass radius of 47 Tuc, so following H05, it appears that the sources in the central 2 arcmin$^2$ are dominated by cluster members, while those in the remainder of the field are not primarily cluster members. As the remainder of the field contains 26 new sources, we estimated (assuming that new non-cluster members in the central region are added at the same rate) that of order 35 new detections are not cluster members, while roughly 46 new cluster members have been added.


\begin{figure}
	\includegraphics[width=\columnwidth]{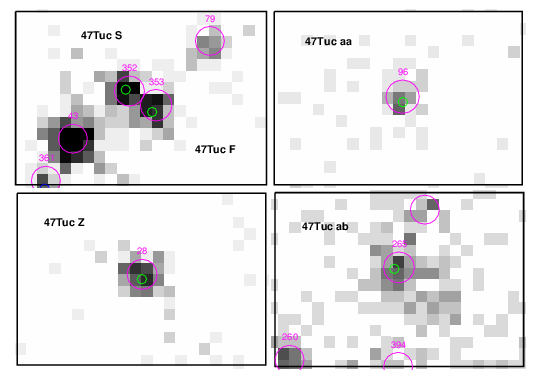}
    \caption{The radio-identified spatial positions of the MSPs studied in this work are shown in green. Their identified X-ray counterparts are indicated in magenta. The background image is a combined 0.5-6 keV image of all \textit{Chandra} ACIS observations of 47 Tuc, binned to half a pixel.}
    \label{fig:MSPpos}
\end{figure}

\begin{table*}
	\caption{X-ray spectral properties of MSPs.}
    \adjustbox{max width=\textwidth}{
	\begin{tabular}{cccccccc}
		\hline
		MSP&Model&kT&R$_{eff}$&L$_{X}$ (0.5-6 keV)&Goodness&Statistic&DoF \\
		& &(keV)&(km)&(10$^{30}$ ergs s$^{-1}$)& (\%)&(C/$\chi^2$) \\
		\hline
	    aa&BBODYRAD&0.20$^{+0.06}_{-0.02}$&0.07$^{+0.06}_{-0.07}$&0.7$^{+0.3}_{-0.2}$&2.88&11.91&9\\ 
		 &NSATMOS&0.14$^{+0.09}_{-0.06}$&0.20$^{+0.53}_{-0.20}$&0.9$^{+0.7}_{-0.8}$&5.71&10.45&9\\
	    ab&BBODYRAD&0.21$^{+0.05}_{-0.04}$&0.09$^{+0.07}_{-0.03}$&1.6$^{+0.3}_{-0.4}$& (10.1) & 14.65 & 9 \\ 
		 &NSATMOS&0.11$^{+0.04}_{-0.03}$&0.45$^{+0.58}_{-0.45}$&1.5$^{+4.8}_{-1.2}$& (18.2) & 12.58 & 9 \\ 
		 Z&BBODYRAD&0.20$^{+0.02}_{-0.02}$&0.14$^{+0.03}_{-0.03}$&2.8$^{+0.4}_{-0.4}$&18.27&29.69&25\\ 
		 &NSATMOS&0.11$^{+0.02}_{-0.02}$&0.68$^{+0.38}_{-0.22}$&2.6$^{+3.1}_{-1.6}$&18.48&24.61&25\\
		S&BBODYRAD&0.18$^{+0.02}_{-0.02}$&0.20$^{+0.06}_{-0.06}$&3.1$^{+0.5}_{-0.5}$& (9.3) & 14.92 & 11 \\ 
		 &NSATMOS&0.09$^{+0.02}_{-0.02}$&1.12$^{+0.74}_{-0.47}$&2.8$^{+5.4}_{-1.8}$& (10.7) & 14.45 & 11 \\
		 F&BBODYRAD&0.19$^{+0.04}_{-0.03}$&0.14$^{+0.06}_{-0.05}$&1.8$^{+0.8}_{-0.3}$&7.57 & 15.27 & 12 \\ 
		 &NSATMOS&0.10$^{+0.03}_{-0.03}$&0.75$^{+0.75}_{-0.75}$&2.1$^{+6.0}_{-1.6}$&8.10  & 12.76  & 12 \\
		\hline
	\end{tabular}\label{table:4}
	}
\ \\\textit{Note}- The BBODYRAD and NSATMOS models used CFLUX and the FLUX command respectively, to determine the unabsorbed luminosity. MSPs ab \& S were fit with the $\chi^2$ statistic (and the null hypothesis probability is quoted rather than the C-stat ``goodness'' statistic), the rest were fit with the C-statistic.
\end{table*}

\begin{figure}
    \centering
	\includegraphics[width=\columnwidth]{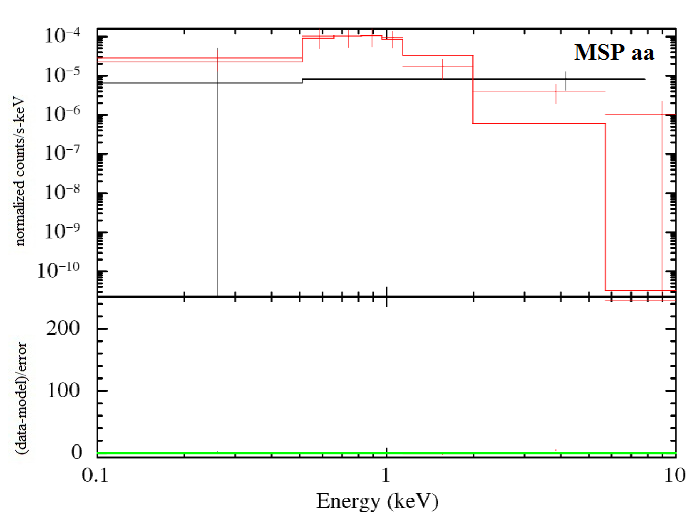}
	\includegraphics[width=\columnwidth]{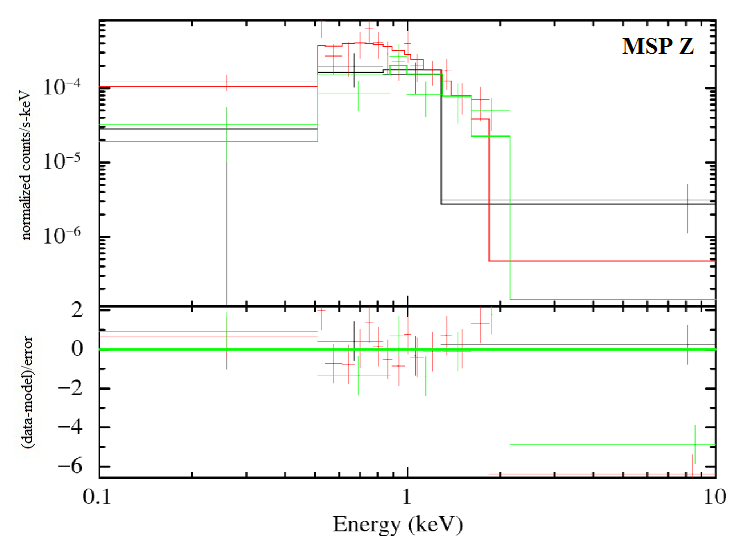}
    \caption{NSATMOS spectral fits to the X-ray counterparts of MSPs 47 Tuc aa and 47 Tuc Z. The spectra were modelled with an absorbed neutron star hydrogen atmosphere (top panels are data and model, lower panels are residuals). The data and spectral fit are shown in black, red and green for the 2000, 2002 and 2014-15 observations respectively, all of which were fit simultaneously.}
    \label{fig:5}
\end{figure}

\section{Spectral Analysis} 
\label{sec:3}
In this paper, we concentrated only on spectral analysis of the newly identified X-ray counterparts to MSPs. Spectra were obtained for the X-ray counterparts of MSPs 47 Tuc F, S, Z, aa and ab separately for the 2000, 2002 and 2014-15 observations using AE (2014-15 observations were not used to fit the X-ray spectra of 47 Tuc aa). Their positions are shown in Fig.~\ref{fig:MSPpos}. For the MSPs 47 Tuc F, Z, and aa, we used the C-statistic, to perform spectral fitting with few photons \citep{Cash 1979}, with bins of 5 counts. On the other hand, the spectra of 47 Tuc S and ab (where we could bin to 10 counts per bin) were fit using the $\chi^2$ statistic.  

Each were first fit to an XSPEC power law, PEGPWRLW, then to the XSPEC blackbody model, BBODYRAD, and finally to a NS hydrogen atmosphere model \citep[NSATMOS;][]{Heinke et al. 2006}, keeping the hydrogen column density frozen to  the cluster value in each case 
\citep[using the TBABS absorption model  with {\it wilm} abundances
;][]{Wilms 2000}, as it could not be reasonably constrained by spectral fits. 
The steep photon indices, over 2.2 (3.02$^{+0.48}_{-0.46}$, 3.09$^{+0.44}_{-0.39}$, 2.89$^{+0.28}_{-0.27}$, 2.29$^{+0.59}_{-0.55}$ and 2.92$^{+0.43}_{-0.38}$ for 47 Tuc F, S, Z, aa and ab respectively), obtained for all five MSPs with power law model fits implied that the spectra are too steep to be produced by typical pulsar magnetospheres, and are likely dominated by thermal emission from the NS surface. 
For the NSATMOS model, the NS mass and radius were fixed to 1.4 $M_{\sun}$ and 11 km, respectively, and the distance to 4.53 kpc, while the normalization was left free (physically interpreted as a variable portion of the surface radiating), as used by e.g. \citet{Bogdanov et al. 2006}. 

\begin{figure}
    \centering
    \includegraphics[width=\columnwidth]{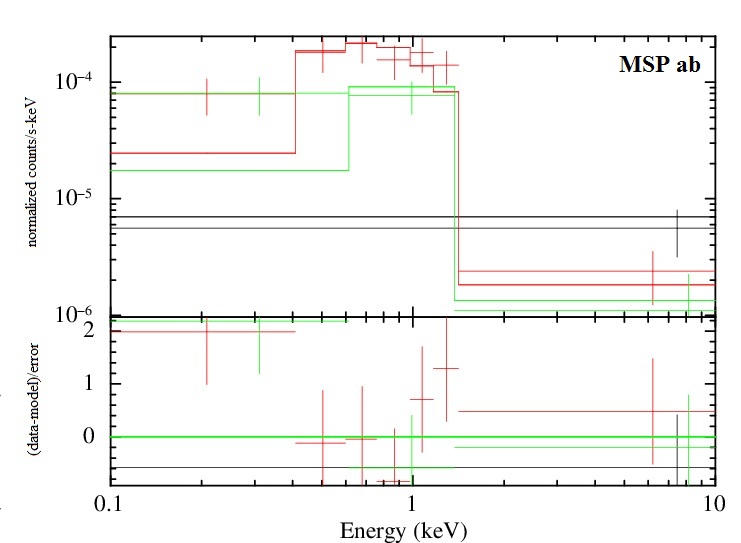}
	\includegraphics[width=\columnwidth]{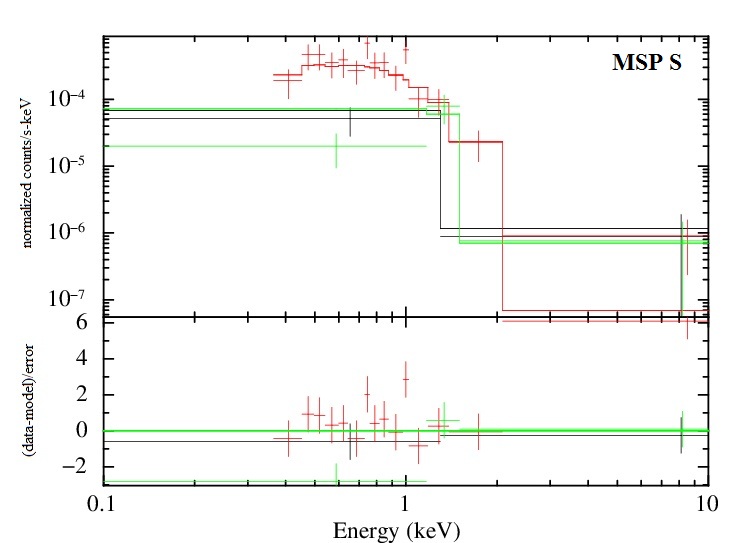}
	\includegraphics[width=\columnwidth]{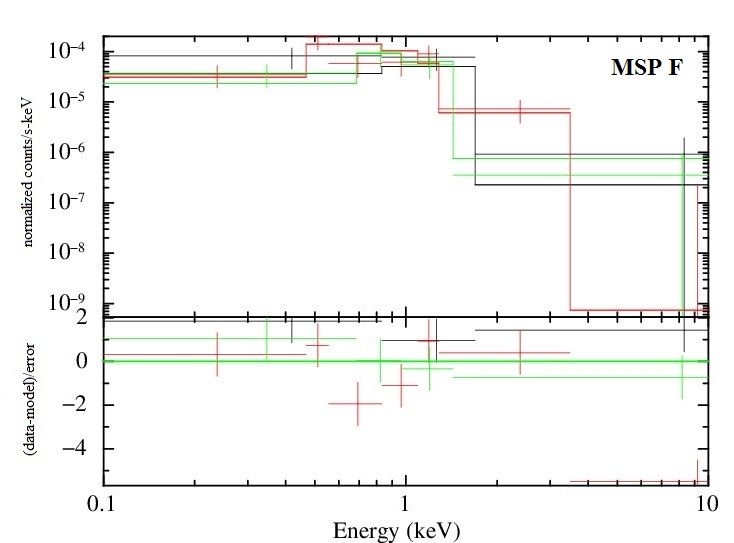}
    \caption{Same as Fig.~\ref{fig:5} but for 47 Tuc ab, 47 Tuc S and 47 Tuc F.}
    \label{fig:5b}
\end{figure}

The results of both model fits for the temperature, radius, and luminosity (0.5-6 keV) are given in Table~\ref{table:4}.  Our fitted temperatures are generally consistent with those found for 47 Tuc MSPs by \citet{Bogdanov et al. 2006}, who found average blackbody/NSATMOS temperatures of 0.18/0.10 keV (range 0.13--0.24/0.07--0.16 keV). Our fitted radii are also generally consistent (\citealt{Bogdanov et al. 2006} found average blackbody/NSATMOS radii of 0.17/0.81 km, ranges of 0.08-0.29/0.28-1.75 km), though 47 Tuc aa has the smallest inferred radius, and the smallest inferred luminosity, of 47 Tuc MSPs. NSATMOS fits are shown in Figs.~\ref{fig:5} and ~\ref{fig:5b}. We used the XSPEC command \lq goodness 1000\rq, which generates 1000 Monte Carlo simulations of the chosen model to see what fraction have a lower fitting statistic than the actual data, to test whether a model is a good fit \citep[e.g.][]{Forestell et al. 2014}. 

\subsection{MSP 47 Tuc aa}
\label{sec:3.1}
47 Tuc aa was discovered by \citet{Pan et al. 2016}, and its timing solution will be presented in Freire \& Ridolfi (2017, in preparation).
We identified W96 as its X-ray counterpart. The possibility of W96 being an MSP was previously suggested by \citet{Edmonds03b}, based on the lack of an optical counterpart, and H05 based on its X-ray properties. It is one of the fainter detected sources, with only 33.5 counts. We did not use the 2014-15 observations to fit the X-ray spectra of 47 Tuc aa, since they lay on the edge of the ACIS subarray. The fit is relatively poor (the fraction of simulated spectra with a worse C statistic than the data is only 5.7\%), apparently due to excess emission at high energies (see Fig. 6), but adding a power-law component (with spectral index between 1 and 2) does not improve the fit. 
The fitted BBODYRAD temperature of 0.20$^{+0.06}_{-0.02}$ keV is consistent with  other MSPs in 47 Tuc \citep{Bogdanov et al. 2006}, but the inferred effective radius  of 0.07$^{+0.06}_{-0.07}$ or 0.20$^{+0.53}_{-0.20}$ km (for blackbody or NSATMOS fits respectively) is the smallest in 47 Tuc. Between $0.5-6$ keV, it has a luminosity of 0.7$^{+0.3}_{-0.2} \times 10^{30}$ ergs s$^{-1}$ (for a blackbody fit, or 0.9$^{+0.7}_{-0.8}\times10^{30}$ erg s$^{-1}$ for the NSATMOS model), making it also the faintest MSP yet identified in 47 Tuc. 

 We also performed fits to 47 Tuc aa assuming emission from the entire surface, in order to calculate constraints on r-mode heating of the interior.  In these fits, we used the model TBABS*(NSATMOS+NSATMOS), where one of the NSATMOS models used a fixed normalization of 1 (corresponding to emission from the entire surface) while the normalization of the other was allowed free (corresponding to emission from the poles).  For these fits only, we explored several neutron star radii; 11, 12, and 12.5 km. Larger stars have weaker constraints on the total luminosity, so we focus on the 12.5 km radius case, selected as roughly the top end of the range of the most plausible neutron star radius estimates (see e.g. \citealt{Lattimer14}).  The 90\% confidence limit on the (unredshifted) surface temperature is $2.80\times10^5$ K, $2.68\times10^5$ K, and $2.62\times10^5$ K for 11, 12, or 12.5 km respectively, leading to (bolometric, calculated as 0.01-10 keV) unredshifted luminosity limits of $3.27\times10^{30}$ erg/s,  $3.43\times10^{30}$ erg/s, or $3.50\times10^{30}$ erg/s, respectively.  

\subsection{MSP 47 Tuc ab}
\label{sec:3.2}
47 Tuc ab was recently discovered by \citet{Pan et al. 2016} who published its timing solution, this was recently updated by \citet{Freire et al. 2017}.  We identify W265 as its X-ray counterpart, although, as noted by H05, W265 appears to consist of multiple confused sources.  W265 is well-fit by blackbody or NSATMOS models with inferred temperature and radius consistent with other MSPs in 47 Tuc, suggesting that the observed X-ray emission at this location is indeed mostly due to 47 Tuc ab.
Its $0.5-6$ keV (blackbody fit) luminosity of 1.6$^{+0.3}_{-0.4} \times 10^{30}$ ergs s$^{-1}$ makes it the second-most X-ray faint MSP identified in 47 Tuc, after 47 Tuc aa. 

\subsection{MSP 47 Tuc Z}
\label{sec:3.3}
\citet{Freire et al. 2017} have found a timing solution for MSP 47 Tuc Z.  We identify W28 as its X-ray counterpart. \citet{Edmonds03b} and H05 both previously suggested that W28 was a possible MSP. The spectral fit to blackbody or NSATMOS models is good, and the inferred parameters are typical of MSPs in 47 Tuc.

\subsection{MSP 47 Tuc S}
\label{sec:3.4}
X-ray emission from MSPs 47 Tuc S and 47 Tuc F was previously studied together, as they had not been resolved \citep{Bogdanov et al. 2006}. With the increased resolution in this work, however, we were able to analyse them separately, identifying W352 as the X-ray counterpart of MSP S. The positional discrepancy, 0.25$''$ in RA and 0.03$''$ in DEC from its known radio position, is slightly larger than typical, which is likely due to crowding. The spectral fits to blackbody or NSATMOS models are good, and the inferred parameters are typical of MSPs in 47 Tuc.

\subsection{MSP 47 Tuc F}
\label{sec:3.5}
We identify W353 as the X-ray counterpart to 47 Tuc F, at 0.27$''$ in RA and 0.15$''$ in DEC from its known radio position. As for 47 Tuc S, the positional discrepancy is larger than average, which we attribute to crowding in this region. The spectral fits to blackbody or NSATMOS models are acceptable, and the inferred parameters are typical of 47 Tuc MSPs. We note that the temperatures for 47 Tuc F and 47 Tuc S individually are consistent with the average temperature for the merged source quoted by \citet{Bogdanov et al. 2006}.

\section{Discussion}
\label{sec:4}
\subsection{Source catalog}
By merging all observations taken with the ACIS instrument of the \textit{Chandra} X-ray Observatory, using the EDSER algorithm, and overbinning the final image, we were able to obtain significantly higher resolution. This, in conjunction with detection algorithms like \textit{wavdetect} and \textit{pwdetect}, and algorithms and tools from AE, allowed us to identify 81 new sources in 47 Tuc, over half of which are present in and near the crowded core region. We thus find a total of 370 sources within the half-mass radius of 47 Tuc. While the newly detected sources in the outskirts of 47 Tuc are probably dominated by background extragalactic sources (considering their relatively uniform spatial distribution), those near the crowded core are likely to be a mixture of ABs and CVs in the cluster, almost certainly dominated by ABs  (see the luminosity functions in Fig. 13 of H05). There may also be a few MSPs and quiescent LMXBs, although new MSPs are likely to be among sources that are newly resolved (e.g. 47 Tuc F and S), rather than sources at the detection limit, based on the known X-ray luminosity function of globular cluster MSPs (see H05).  {MSPs generally have luminosities (0.5-2.5 keV) in the range $10^{30}-10^{31}$ erg s$^{-1}$, while quiescent LMXBs range from $\sim 10^{31}-10^{33}$ erg s$^{-1}$. H05 was already complete for uncrowded sources with luminosities above $8\times 10^{29}$ ergs s$^{-1}$ and this work only pushes the detection limit marginally. So the increased sources we detect are largely due to improvement in the angular resolution and detection efficiency.}\\

\begin{figure}
	\includegraphics[width=\columnwidth]{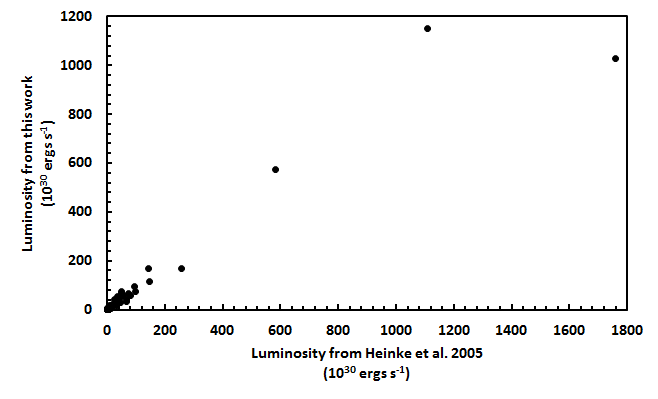}
	\includegraphics[width=\columnwidth]{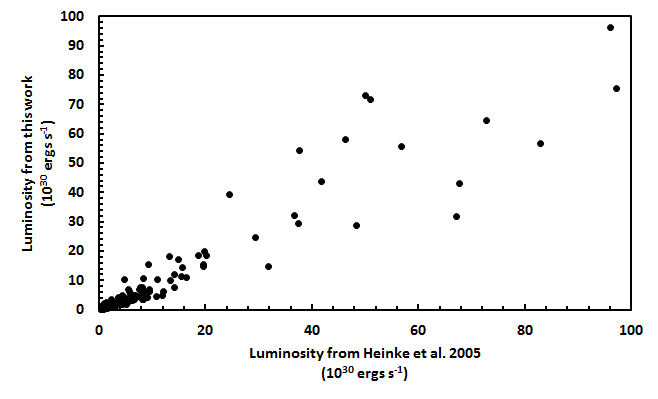}
    \caption{Comparison of luminosities in the 0.5-6 keV band derived by this work with that obtained by H05. The top panel shows all the sources identified in both works while the bottom panel shows those sources with luminosities less than 100$\times10^{30}$ ergs s$^{-1}$.}
    \label{fig:6}
\end{figure}

The luminosities for these sources, computed from the photometry as explained in \S \ref{sec:2.3}, do not always match those obtained by H05 as illustrated in Fig.~\ref{fig:6}. Sources with higher counts in the 2-6 keV range, notably W42 (X9), seem to show an increased luminosity in this work, whereas sources with lesser counts in this hard band, notably W46 (X7), seem to show a decreased luminosity (though spectral fitting shows that W46/X7 has demonstrated that it is not actually variable; \citealt{Bogdanov et al. 2016}). A likely explanation is the reduced sensitivity of the ACIS-S detector in the lower energy bands, which combined with a single assumed spectral model can produce such apparent variations. (For example, the expected 0.5-6 keV countrate for an on-axis source described by a power-law with photon index 1.8 and the cluster $N_H$ decreases by 28\% from the 2002 to 2014 ACIS-S observations; for a 2 keV VMEKAL model, the decrease is 30\%.) It is to be noted that the individual luminosities of these sources must be obtained more accurately by fitting different spectral models suitable for different objects.  {W42 (X9) has been comprehensively studied by \citet{Bahramian et al. 2017} who confirm this source as an ultracompact X-ray binary with a C/O white dwarf donor and a possible black hole primary.}

\subsection{MSP polar caps}
From the spectral fits for the recently discovered MSPs 47 Tuc aa, 47 Tuc ab and 47 Tuc Z, and freshly resolved MSPs 47 Tuc S and 47 Tuc F, we obtained their temperatures and luminosities, which were found to be consistent with those of most other MSPs in 47 Tuc, analysed by \citet{Bogdanov et al. 2006}. In comparison to the BBODYRAD model, the NSATMOS hydrogen atmosphere model gives lower estimates of the temperatures (Table~\ref{table:2}), while the unabsorbed luminosity estimates from the two models are consistent.  
Since the NS surface is almost certainly covered with a hydrogen atmosphere in these MSPs \citep{Zavlin Pavlov 1998,Bogdanov Rybicki Grindlay 2007}, the (larger) NSATMOS model estimates of the emitting radii are more realistic. As discussed by \citet{Bogdanov et al. 2006}, these estimates of the emitting radius will vary from the true polar cap sizes, as high-quality spectra of the nearby MSP PSR J0437$-$4715 show that at least two, probably three, thermal components are required \citep{Zavlin02,Bogdanov 2013,Guillot16}, likely from different parts of the polar caps.

\begin{figure}
    \centering
	\includegraphics[width=\columnwidth]{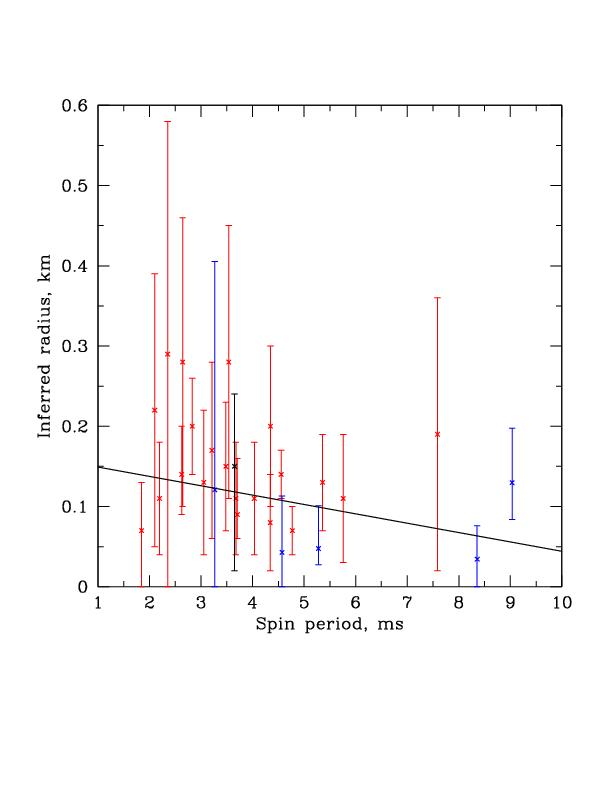}
    \caption{Fitted (BBODYRAD) polar cap radius against spin period for MSPs in 47 Tuc (red), NGC 6397 (black) and NGC 6752 (blue). The best fitting power law is indicated, with a best fitting slope $-0.41 \pm 0.27$, consistent with the predicted index of $-0.5$. }
    \label{fig:spin}
\end{figure}

It has long been predicted that the size of the polar cap region of a radio pulsar, $R_{\rm pc}=(2 \pi R_{\rm NS}/(cP))^{1/2}R_{\rm NS}$ \citep[e.g.][]{Lyne Graham-Smith}, where $R_{\rm NS}$ is the radius of the NS having period P,  depends inversely on the spin period. It thus follows that MSPs in 47 Tuc, having shorter periods on average than those in NGC 6752, should have larger polar caps (given similar luminosities). By fitting the effective radii measurements of MSPs in 47 Tuc, NGC 6752 and NGC 6397, with a power law in spin period, \citet{Forestell et al. 2014} found a best-fitting index of $-0.65 \pm 0.40$ ($1\sigma$ error bars), consistent with the predicted index of $-0.5$. We plotted the spin periods against the inferred MSP effective radii (from BBODYRAD) for all the MSPs in 47 Tuc, NGC 6752 and NGC 6397, adding six new pulsars to 47 Tuc (those studied here, and 47 Tuc X from \citealt{Ridolfi16}; Fig.~\ref{fig:spin}). 
Fitting the effective radii measurements with a power law in spin period, we found a best-fitting index of $-0.41 \pm 0.27$ ($1\sigma$ error bars). This slightly increases the evidence for a correlation by reducing the $1\sigma$ uncertainty, and is still consistent with the theoretically predicted index of $-0.5$.  Dispersion into this correlation is expected from the (unknown) differences in geometries of the pulsars, and by variations in the strength of unmodelled non-thermal radiation. 

\section{Upper limit for surface temperature of MSP 47 Tuc aa and constraints on r-modes}

An intriguing consequence of the low X-ray luminosity of 47 Tuc aa is the constraint it allows us to impose upon internal heating in rapidly spinning neutron stars. MSPs should be in thermal balance, because their thermal evolution timescale is much shorter than their spin down time-scale. Thus the total heating power should be compensated by cooling, which depends on the MSP temperature.

The surface temperature of 47 Tuc aa is low enough (see \S \ref{sec:3.1}) to exclude strong neutrino emission from the bulk of the star (see e.g. \citealt{cgk17} for more detailed discussion). Thus, the total cooling power of this source can be estimated almost directly from observations: it equals the thermal emission from the entire surface, excluding  hot spots (i.e. from the component with a fixed normalization of 1 in \S \ref{sec:3.1}), %
\footnote{Here we use the least constraining limit on the bolometric luminosity from the entire surface of 47 Tuc aa from \S \ref{sec:3.1}, $L\le 3.50\times10^{30}$ erg/s for an assumed radius of 12.5 km. }
\begin{equation}
L_\mathrm{cool}\approx L\le 3.5\times 10^{30}\,\mathrm{erg\,s}^{-1}, \label{Lcool}
\end{equation}
providing the same constraint to the total heating power, which can be presented as a sum of internal heating mechanisms (namely,  superfluid vortex creep \citealt{Alpar_etal84},
rotochemical heating \citealt{Reisenegger95}, 
rotation-induced deep crustal heating \citealt{gkr15}), and possible heating produced by the dissipation of unstable oscillation modes (e.g., r-modes, 
\citealt{ajk02,rb03,cgk17,Schwenzer_etal_Xray}).

We start from the internal heating. Its power $Q$ is generally proportional to the (intrinsic) spin down rate (\citealt{gr10,gkr15})
which can be bounded as $\dot \nu\ge -5\times10^{-15}\, \mathrm{s}^{-2}$ on the basis of the accurate timing solution (Freire \& Ridolfi 2017, in preparation) and accounting for the strongest possible negative acceleration in the cluster gravitational field (\citealt{Freire et al. 2017}). 
The typical spin-down power of other MSPs with X-ray luminosity $\sim 10^{30}$\,erg\,s$^{-1}$ (see, e.g., Fig. 8 in \citealt{Forestell et al. 2014}) suggests that real spin-down rate of  47Tuc aa is slower, but for estimates below we apply robust bound from the radio observations. Following \cite{cgk17}, to estimate the internal heating power  we appeal to the rotation-induced deep crustal heating (\citealt{gkr15}), which is associated with the same physics as deep crustal heating in accreting NSs (\citealt{bbr98}). Namely, compression of the accreted material (due to subsequent accretion for accreting NSs, or due to spin-down in case of MSPs)  leads to nuclear reactions in the crust and production of heat. 
This mechanism does not depend on the uncertain parameters of superfluid transition in the star and thus it is rather robust. For the parameters of 47 Tuc aa it gives
\begin{eqnarray}
	Q&\gtrsim& Q_\mathrm{DCH}\approx
	1.8\times 10^{30} \frac{\mathrm{erg}}{\mathrm s}
	\left(\frac{R}{12.5\,\mathrm{km}}\right)^7\,
	\left(\frac{M}{1.4\, M_\odot}\right)^{-2}\,
	\nonumber \\
	&\times& \frac{\nu}{542\,\mathrm{Hz}}\frac{\left|\dot\nu\right|}{5\times 10^{-15}\,\mathrm{s}^{-2}}
	\sum_i \frac{P_i}{10^{31}\, \mathrm{erg\,\, cm}^{-3}}\frac{ q_i}{\mathrm{MeV}},
	\label{QDCH}
\end{eqnarray}

where Eq.\ (7) from \cite{gkr15} was applied (parameter $a$ taken to be $a=0.5$). Here $\left|\dot\nu\right|=-\dot\nu$ is absolute value of the intrinsic spin-down rate; $i$ enumerates the different reactions in the crust; $P_i$, and $q_i$ are, respectively, the threshold  pressure and energy production for each reaction. For all models of the accreted crust discussed by \cite{hz08} $\sum_i P_i q_i$ lies in the range $(0.9-1.5)\times 10^{31} \,\mathrm{erg\, MeV\, cm}^{-3}$, making $Q_\mathrm{DCH}$ well constrained, and close to the  observational upper bound (\ref{Lcool}). Note, $Q_\mathrm{DCH}$ is strongly increasing with increase of $R$ ($Q_\mathrm{DCH}\propto R^7$), thus for larger radius or if other internal heating mechanisms are  competitive with  $Q_\mathrm{DCH}$, the total heating power can exceed constraint (\ref{Lcool}) for $\dot \nu=-5\times10^{-15}\, \mathrm{s}^{-2}$, suggesting thus that the intrinsic spin-down rate of MSP 47 Tuc aa should be lower. However, we leave detailed analysis of such constraints beyond the scope of the paper because they are model dependent.

\begin{figure*}
	\includegraphics[width=16cm]{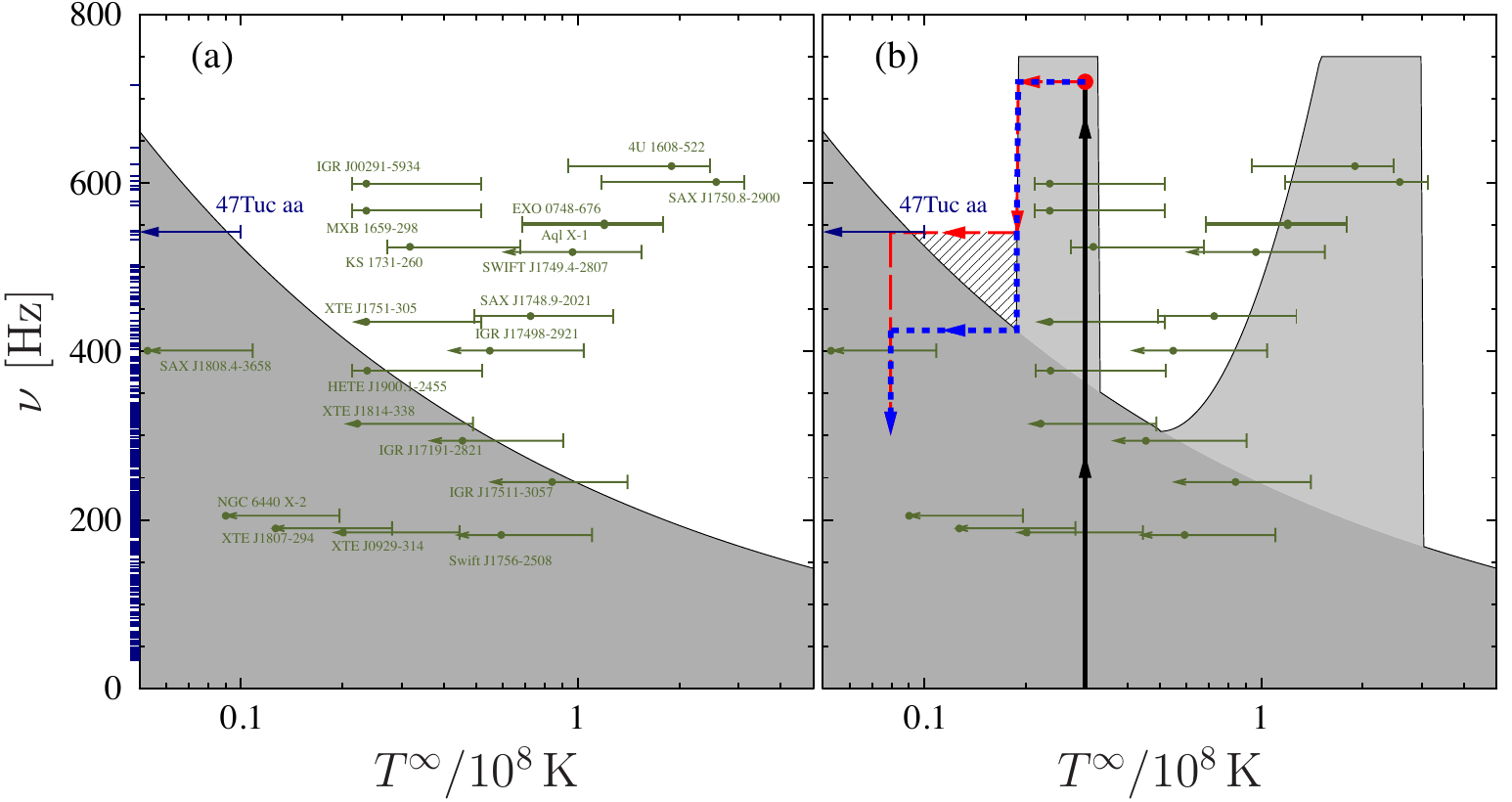}
	\caption{Examples of r-mode instability windows in the standard  model [panel (a)] and the
		minimally constrained model  [panel (b), see text for details].
		The stability region is shaded in grey;
		in the white region the r-mode is unstable.
		Temperatures and frequencies
		of NSs observed in LMXBs are shown by filled circles,
		while error bars show uncertainties due to the unknown NS envelope composition. 
		The upper limit for the internal temperature of MSP 47 Tuc aa is also marked.
		Ticks on the left side of $\nu$ axis show measured MSP frequencies.		
	}
	\label{Fig_rmode}
\end{figure*}

The heating by unstable modes can take place in rapidly rotating neutron stars  due to the Chandrasekhar-Friedman-Schutz (CFS;
\citealt{chandrasekhar70a,fs78a,fs78b,andersson98,fm98}) instability, driven by emission of gravitational waves.
In the absence of dissipation, the CFS instability takes place at an arbitrary rotation rate and results in the exponential growth of a certain class of oscillation modes. The most unstable among them are r-modes (predominantly toroidal modes, which are similar to Rossby waves and controlled by the Coriolis force).
Dissipation suppresses the instability only up to a threshold spin frequency (which depends on the internal NS temperature; see Fig.\ \ref{Fig_rmode} and e.g.\ \citealt{haskell15} for a recent review).  
The standard model  of r-mode instability (suggested by \citealt*{lom98}; \citealt{olcsva98}, see also \citealt*{gck14b} for discussion of recent microphysical updates) assumes a hadronic composition of the NS core and dissipation by shear and bulk viscosity. It stabilizes the NS in the grey region in Fig.\ \ref{Fig_rmode}(a), the unstable (white) region is referred to as the `instability window'. 

As shown by \cite{hah11,hdh12}, the observations of transiently accreting neutron stars (their spin frequencies and internal temperatures are shown in Fig.\ \ref{Fig_rmode} by dots with error bars associated with uncertainty in thermal insulating envelope composition; data are taken from \citealt*{gck14a,cgk17}) reveal the inadequacy of the standard model: additional dissipation of r-modes is required to stabilize many observed NSs [i.e., all NSs in the white region of \ref{Fig_rmode}(a)]. \cite{cgk17} analyze the formation of MSPs via the recycling scenario (\citealt*{bkk76}; \citealt{acrs82,Bhattacharya van den Heuvel 1991, Papitto et al. 2013})
and suggest `minimal' constraints to the instability windows, which allow them to explain the observations of transiently accreting neutron stars, and the formation of high-frequency MSPs, simultaneously.  The corresponding `minimally constrained' instability window is shown in Fig. \ref{Fig_rmode}(b). As indicated by \cite{cgk17}, strong upper bounds on thermal emission from the  entire surface of rapidly rotating MSPs can lead to even stronger constraints, and here we apply the upper limit (\ref{Lcool}) to this aim. 

The first constraint comes from the thermal equilibrium in the current state of the MSP. As discussed by \cite{cgk17,Schwenzer_etal_Xray,Mahmoodifar17}, the upper limit for the MSP surface temperature gives an upper bound to the heating by r-modes, and thus to the r-mode amplitude. Substituting the upper limit on the surface temperature from \S \ref{sec:3.1} into Eq.\ (12) from \cite{cgk17}, we get the constraint
\begin{equation}
\alpha\lesssim 2.5\times 10^{-9}
\label{alpha_47Tuc}
\end{equation}
for the r-mode amplitude (defined as in \citealt{lom98}) in MSP 47 Tuc aa. The numerical value corresponds to a radius $R=11$~km, which gives the least constraining bound (assuming $R=12$~km we come to $\alpha\lesssim 2\times 10^{-9}$).
It is worthwhile to note, that Eq.\ (\ref{alpha_47Tuc}) is the strongest constraint available for MSPs (\citealt{Schwenzer_etal_Xray,cgk17,Mahmoodifar17})%
\footnote{\cite{Schwenzer_etal_Xray} suggest a comparable constraint for  PSR J1023+0038 ($\nu=592.4$~Hz), but in this paper the surface temperature of  PSR  J1023+0038 was taken to be  $T^\infty_\mathrm{eff}\lesssim 3\times 10^5$\,K (see pulsar at $\nu=592$~Hz in Figure 1 of that paper), without detailed spectral fitting.  However, such a strong upper limit is not justified for J1023+0038 (e.g., the joint fit of X-ray spectra by \cite{Bogdanov_etal11} suggests a higher surface temperature $T^\infty_\mathrm{eff}\sim4\times 10^5$~K, leading thus to weaker constraints on the r-mode amplitude).}
 and accreting neutron stars (\citealt{ms13}).
As long as non-linear saturation of the r-mode instability predicted by state-of-art models  (\citealt*{arras_et_al_03,btw07,bw13,hga14}) takes place at much larger amplitudes, the most natural explanation of  the bound (\ref{alpha_47Tuc}) is that MSP 47 Tuc aa is stable (at least with respect to CFS instability of r-modes). Indeed, the upper limit for the surface temperature of MSP 47 Tuc aa allows us to constrain the redshifted internal temperature as $T^\infty\le 10^7$~K even for the iron thermally insulating envelope model by \citealt{pcy97} [a layer of light (accreted) elements  with mass $\Delta M>10^{-13}M_\odot$ reduce this  constraint to $T^\infty\le 5\times 10^6$~K], which is almost enough to guarantee stability of 47 Tuc aa in the standard r-mode instability model [see Fig.\ \ref{Fig_rmode}, panel (a)].

The second constraint, developed by \cite{cgk17}, appeals not only to  the current state of this MSP, but also to its formation. Namely, it comes from the requirement that the r-mode instability does not prevent cooling of the MSP to the observed temperature, after the end of accretion (and spinup) during the LMXB stage. \cite{cgk17} apply  $T^\infty \lesssim 2\times 10^7$\,K as a fiducial upper limit for internal temperatures of MSPs and demonstrate that it leads to a `minimally' constrained instability window [see Fig.\ \ref{Fig_rmode} (b)]. Since MSP 47 Tuc aa has a lower internal temperature, it puts a stronger constraint on the shape of the instability window. 

For the minimally constrained instability window from \cite{cgk17}, the evolution of the MSP progenitor during the LMXB stage is unaffected by the r-mode instability, and the evolutionary trajectory is shown schematically by the thick solid line in Fig.\ \ref{Fig_rmode} (b). After the end of accretion, the newly born MSP  loses accretion-induced heating, cools down and evolves along the low temperature boundary of the stability peak, where r-modes force spin-down of the MSP and keep it heated (dotted [blue] line in the plot). It is easy to see that this line cannot explain the temperature and spin of MSP 47 Tuc aa (namely, this path predicts its internal temperature to be $\sim 2\times 10^7$~K, corresponding to $T_\mathrm{eff}^\infty\sim 6\times 10^5$\,K (an accreted  thermal insulating envelope is assumed), which is well above the upper limit obtained in \S \ref{sec:3.1}). To allow MSP 47 Tuc aa to cool down enough to be in agreement with observations, the r-mode instability should be suppressed at $(1-2)\times 10^7$~K at least for the frequency of this pulsar ($\nu\sim 542$~Hz, \citealt{Pan et al. 2016}). 
Assuming that there are no isolated regions of r-mode instability, r-modes must be stable in the  shaded region in Fig.\ \ref{Fig_rmode} (b).  In this case, the evolution of MSPs can follow the dashed (red) line in this figure, which is in agreement with observations of MSP 47 Tuc aa. 

\section{Conclusions}

 {We combined 180 ks of new Chandra ACIS data on 47 Tuc with 370 ks of archival data, and used improved algorithms to generate a new source catalog, finding 81 new sources for a total of 370 within the half-mass region. Roughly half of the new sources are likely associated with the cluster, and half are background AGN. We resolved the X-ray emission from MSPs 47 Tuc F and 47 Tuc S, and use recent pulsar timing solutions to identify X-ray emission from the MSPs 47 Tuc aa, 47 Tuc ab, and 47 Tuc Z.  In general, their X-ray emission is consistent with that of other MSPs in 47 Tuc, though 47 Tuc aa is the X-ray faintest MSP yet measured in 47 Tuc. Comparing the fitted blackbody radii of millisecond pulsar polar caps with their spin rates, we find modest evidence for the expected anticorrelation. Finally, we use our upper limit on the temperature of the surface of the fast-spinning (542 Hz) MSP 47 Tuc aa to constrain the heating of this neutron star by r-modes. We find a constraint on the amplitude of r-modes in 47 Tuc aa of $\lesssim2.5\times10^{-9}$, the most constraining yet for MSPs.  We also use the temperature and rotation frequency of 47 Tuc aa to place a constraint on the shape of the r-mode instability window in neutron stars.}

\section*{Acknowledgements}
S. Bhattacharya acknowledges support from MITACS for sponsoring his stay at the University of Alberta. COH acknowledges support from an NSERC Discovery Grant, and an NSERC Discovery Accelerator Supplement. AR and PCCF gratefully acknowledge financial support by the European Research Council for the ERC Starting grant BEACON under contract no. 279702. AR is a member of the International Max Planck research school for Astronomy and Astrophysics at the Universities of Bonn and Cologne and acknowledges partial support through the Bonn-Cologne Graduate School of Physics and Astronomy. This work was funded in part by NASA Chandra grant GO4-15029A awarded through Columbia University and issued by the Chandra X-ray Observatory Center (CXC), which is operated by the Smithsonian Astrophysical Observatory for and on behalf of NASA under contract NAS803060. This research has made use of the NASA Astrophysics Data System (ADS) and software provided by the CXC in the application package CIAO. \textit{pwdetect} has been developed by scientists at Osservatorio Astronomico di Palermo G. S. Vaiana thanks to Italian CNAA and MURST(COFIN) grants.

\bibliographystyle{mnras}

\bsp	
\label{lastpage}

\end{document}